\documentclass[preprint,letterpaper]{revtex4}
\usepackage{graphicx}
\def \uU {{\underline U}^{rot}}
\def \ta {\theta_a}
\def \tb {\theta_b}
\def \dab {\delta_{\alpha \beta }}
\def \ab {\alpha \beta }
\def \ds {\displaystyle}
\def \ua {\uparrow }
\def \da {\downarrow }
\def \bc {\bar c}
\def \bS {\bar S}
\def \bas {\bar s}
\def \te {\tilde \epsilon}
\def \tm {\tilde \mu}
\def \bs {{\bf \sigma }}
\def \dsw {D_{{\rm sw}}}
\def \dsd {\Lambda }
\def \eke {K}
\def \wk {w_{{\bf k}}}
\def \TC {T_{\rm C}}
\def \JH {J_{\rm H}}
\def \vS {{\bf S}}
\def \vB {{\bf B}}
\def \vx {\hat {\bf x}}

\def \vz {\hat {\bf z}}
\def \vQ {{\bf Q}}

\def \vs {{\bf s}}

\def \vk {{\bf k}}
\def \vq {{\bf q}}
\def \sgn {{\rm sgn}}
\begin{document}

\title{Double Exchange in a Magnetically Frustrated System}
\author{R.S. Fishman}
\affiliation{Condensed Matter Sciences Division, Oak Ridge National Laboratory, Oak Ridge, TN 37831-6032}

\begin{abstract}

This work examines the magnetic order and spin dynamics of a double-exchange model 
with competing ferromagnetic and antiferromagnetic Heisenberg interactions between the local moments.  
The Heisenberg interactions are periodically arranged in a Villain configuration
in two dimensions with nearest-neighbor, ferromagnetic coupling $J$ and antiferromagnetic coupling $-\eta J$.  
This model is solved at zero temperature by performing a $1/\sqrt{S}$ expansion in the rotated reference 
frame of each local moment.  When $\eta $ exceeds a critical value, the ground state is a magnetically 
frustrated, canted antiferromagnet.  With increasing hopping energy $t$ or magnetic field $B$, the 
local moments become aligned and the ferromagnetic phase is stabilized above critical values of $t$ or $B$.  
In the canted phase, a charge-density wave forms because the electrons prefer to sit on lines of sites that 
are coupled ferromagnetically.  Due to a change in the topology of the Fermi surface from closed to open, phase 
separation occurs in a narrow range of parameters in the canted phase.  In zero field, the long-wavelength 
spin waves are isotropic in the region of phase separation.  Whereas the average spin-wave stiffness in 
the canted phase increases with $t$ or $\eta $, it exhibits a more complicated dependence on field.  
This work strongly suggests that the jump in the spin-wave stiffness observed in Pr$_{1-x}$Ca$_x$MnO$_3$ with 
$0.3 \le x \le 0.4$ at a field of 3 T is caused by the delocalization of the electrons rather than by the alignment 
of the antiferromagnetic regions. 

\end{abstract}

\maketitle

\section{Introduction}

The persistence of antiferromagnetic (AFM) short-range order in the ferromagnetic (FM) phase of the 
manganites has been recognized for many years \cite{dag:01}.  In metallic manganites like La$_{0.7}$Ca$_{0.3}$MnO$_3$ 
that contain a preponderance of AFM-coupled polaronic regions \cite{ada:00,koo:01},  the Curie temperature 
$\TC $ is suppressed but the magnetoresistance is strongly enhanced.  Close to but below $\TC$, the 
spin dynamics of La$_{0.7}$Ca$_{0.3}$MnO$_3$ contains both a propagating spin-wave (SW) branch from the 
FM regions and a diffusive component from polaronic regions with suppressed FM interactions \cite{che:03}.  
The low-temperature insulating phase of the manganite Pr$_{1-x}$Ca$_x$MnO$_3$ with $0.3 \le x \le 0.4$ was 
originally believed \cite{jir:85,yos:95,oki:99} to be a canted AFM (CAF) but may actually 
contain regions with both FM and AFM short-range order \cite{deac:01,rad:01,har:01,fer:02,sim:02,mer:03}.  
When an applied field $B$ exceeds about 3 T, the resistivity drops by several orders of magnitude \cite{yos:95}, 
the AFM regions shrink \cite{sim:02}, and the SW stiffness $\dsw $ jumps by a factor of 3 \cite{fer:02}.  
Despite the central role played by AFM interactions in the manganites, little is known theoretically 
about how they affect the propagating SW dynamics.  As first shown by Anderson and Hasegawa \cite{and:55}, the effective 
electron-hopping energy between two local moments making a relative angle $\Theta $ is proportional 
to $\cos \Theta /2$ in the limit of large Hund's coupling.  So AFM interactions may suppress the contribution 
of electron-mediated double-exchange (DE) to the SW dynamics \cite{deg:60}.  By aligning the local
moments, a magnetic field or electron hopping will alter the DE contribution to the SW dynamics. 
This paper examines the effects of AFM interactions on the ground-state properties and SW dynamics 
of electrons coupled to the local moments of a generalized Villain model \cite{vil:77,ber:86,gab:89,sas:92}.

The generalized Villain model is one of the simplest periodic models to exhibit
magnetic frustration.  As described in Fig.1(a), the local moments $\vS_i$ are 
coupled by the FM interaction $J$ along the $y$ direction and by either the FM interaction $J$ 
or the AFM interaction $-\eta J$ along the $x$ direction.  The CAF phase is stable
when $\eta $ exceeds the critical value $\eta_c$, which is $1/3$ when $\vB=B\vz = 0$ but increases 
as $B$ increases.  Due to the different environments of the $a$ and $b$ sites, the angle $\tb $ 
at the $b$ sites is always larger than $\ta $ at the $a$ sites, as shown in Fig.1(b).

\begin{figure}
\includegraphics *[scale=0.8]{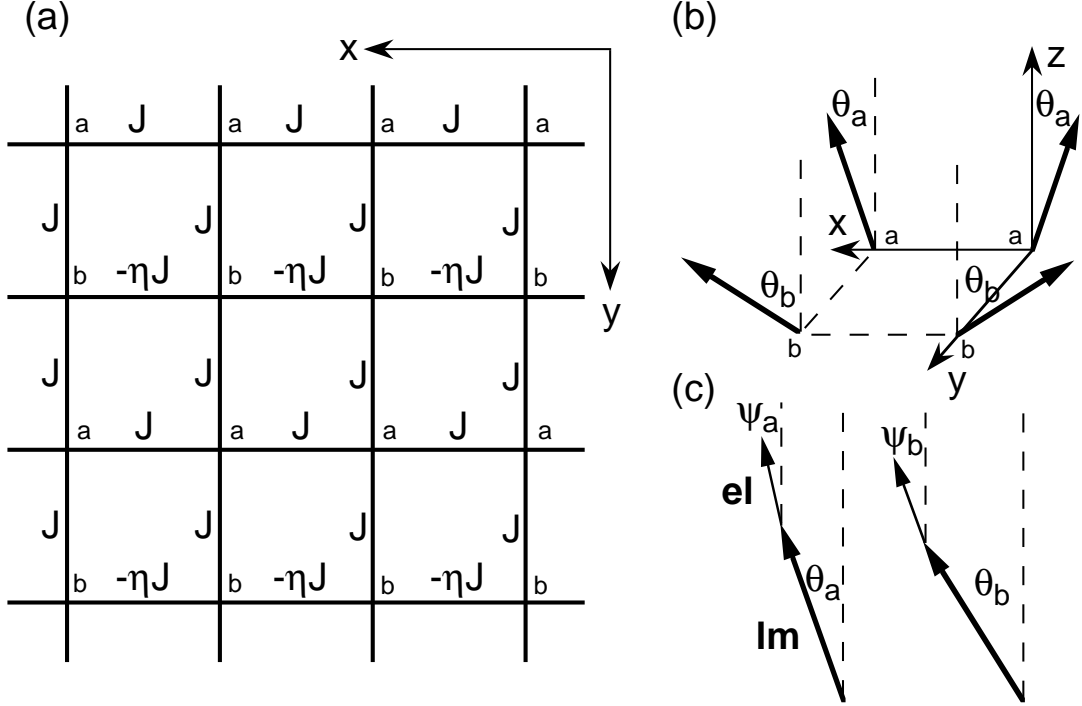}
\caption{
(a)  The generalized Villain model with Heisenberg couplings $J$ or $-\eta J$, (b) the
local moments in the $xz$ plane subtend angles $\theta_a$ and $\theta_b$ with the $z$ axis, 
and (c) the electron spins also lie in the $xz$ plane but subtend angles $\psi_a < \theta_a$
and $\psi_b < \theta_b $ with the $z$ axis.
}
\end{figure}

In the hybrid model considered here, the Heisenberg interactions between the local moments 
are in the Villain configuration while electrons with density $p=1-x$ are FM coupled 
to the local moments by Hund's coupling $\JH$ and hop between neighboring 
sites with energy $t$.  The DEV model (so called because it combines the DE and Villain models) 
provides several advantages as a basis for understanding the effects of AFM interactions and 
non-collinearity on the spin dynamics.  First, except in a narrow range of parameters, a homogeneous 
CAF phase is stable against phase separation when the AFM control parameter $\eta $ exceeds $\eta_c$.
By contrast, the well-studied hybrid model with AFM interactions $-J$ between all neighboring 
local moments phase separates {\it before} the AFM exchange $J$ is large enough to cant the spins 
\cite{yun:98b,gol:98,kag:99,gol:00}.  This phase instability is caused by a ``site-local continuous degeneracy'' 
\cite{gol:98,gol:00,inst} that is absent in the DEV model.  Second, unlike the case in a hybrid model 
with AFM exchange only, the ground state of the DEV model contains a FM component even when $t=0$ and $B=0$.  
So it can be used to evaluate the change in SW stiffness $\dsw $ as the electrons become mobile.  Third, 
because the DEV model contains both FM and AFM Heisenberg interactions, it can be used to study 
insulating manganites like Pr$_{0.66}$Ca$_{0.34}$MnO$_3$, where the AFM interactions are produced
by superexchange and the FM interactions by short-ranged orbital order \cite{miz:00,pol}.  

For simplicity, we have constructed a model that is translationally symmetric in two dimensions.  
Since this model is solved at zero temperature, the qualitative results will be unchanged 
in three dimensions.  More problematically, the AFM interactions are arranged periodically rather than 
in clusters.  In the low-temperature phase of Pr$_{0.67}$Ca$_{0.33}$MnO$_3$, the FM interactions may 
be confined to two-dimensional sheets with widths of roughly $25 \AA $ in a ``red cabbage'' 
structure \cite{sim:02,mer:03}.  Neutron-scattering results \cite{fer:02}, on the other hand,
suggest that the FM clusters in the insulating phase are about $40 \AA $ is diameter.  So for 
wavelengths much longer than $40 \AA $, the SW's will average over the FM and AFM regions.  
Hence, the DEV model will provide qualitatively accurate predictions for the average SW stiffness 
$\dsw^{av}=(\dsw^x+\dsw^y)/2$ defined in the long-wavelength limit.
   
The Hamiltonian of the DEV model is given by
\begin{equation}
\label{ham}
H=-t\sum_{\langle i,j \rangle }\sum_{\alpha }\Bigl( c^{\dagger }_{i\alpha }c^{\, }_{j\alpha }
+c^{\dagger }_{j\alpha }c^{\, }_{i\alpha } \Bigr) -2 \JH \sum_i \vs_i \cdot \vS_i
-\sum_{\langle i,j \rangle }J_{ij}\vS_i \cdot \vS_j -B\sum_i S_{iz} ,
\end{equation}
where $c^{\dagger }_{i\alpha }$ and $c_{i\alpha }$ are the creation and
destruction operators for an electron with spin $\alpha $ at site $i$,
$\vs_i =(1/2) c^{\dagger }_{i\alpha } \bs_{\alpha \beta } c^{\, }_{i\beta }$
is the electronic spin, and $\vS_i$ is the spin of the local moment with magnitude $S$. 
Nearest-neighbor Heisenberg interactions $J_{ij}$ take the values $J$ (FM interaction)
or $-\eta J$ (AFM interaction), as described in
Fig.1(a).  This model is solved at zero temperature by expanding the Hamiltonian in
powers of $1/\sqrt{S}$.  To guarantee that the contributions to the 
SW frequencies from hopping and from the Heisenberg interactions are of the 
same order in $1/\sqrt{S}$, $t$ is considered to be of the same order in $1/\sqrt{S}$ as 
$\JH S$, $JS^2$ and $BS$.  Hence, the dimensionless parameters of our model are $t'=t/JS^2$, 
$\eta $, $B'=B/JS$, and $\JH /JS$.  To lowest order in $1/S$, the magnetic field $B$ only couples 
to the local moments and not to the electrons.  While the theory developed below can be extended 
to treat all values of the Hund's coupling, we shall for simplicity consider the limit of large $\JH S$ 
or in dimensionless terms, $\JH /JS \gg 1$ and $\JH S/t \gg 1$.

This paper is divided into five sections.  The Villain model is discussed in some
detail in Section II, where we provide new results for the SW stiffness. 
The ground-state properties of the DEV model are presented in Section III.  In Section
IV, we evaluate the SW frequencies of the DEV model.  Section V contains a discussion 
and summary.  A short version of this work \cite{fis:04} is in press.

\section{Generalized Villain Model}

This section presents, for the first time, a Holstein-Primakoff expansion for the 
generalized Villain model.  In an equivalent approach, Saslow and Erwin \cite{sas:92}
numerically evaluated the mode frequencies by linearizing the equations of motion for the spins.
However, a formal Holstein-Primakoff expansion is required to lay the foundation for 
the solution of the full DEV model in Section IV.
 
The Hamiltonian of the generalized Villain model is given by Eq.(\ref{ham}) with $t=0$
and $\JH =0$.  The spin dynamics is immensely simplified in the rotated reference frame for each spin:  
$\bar \vS_i =\uU_i \vS_i$, where $\uU_i$ is the unitary rotation matrix for site $i$.
A Holstein-Primakoff expansion is performed within each rotated reference frame:
$\bS_{iz} =S-a_i^{\dagger }a_i$, $\bS_{ix}+i\bS_{iy}=\sqrt{2S}a_i$, and 
$\bS_{ix}-i\bS_{iy} =\sqrt{2S}a_i^{\dagger }$.  
Then the zeroth-order term (in powers of $1/\sqrt{S}$) in $H_v$ can be written as
\begin{equation}
\label{Eh}
E_h=\ds\frac{1}{2}NJS^2\biggl\{ -\cos 2\ta +\eta \cos 2\tb -2\cos (\ta -\tb )
-B'  \Bigl( \cos \ta +\cos \tb \Bigr) \biggr\},
\end{equation}
which is of order $JS^2$.  

Minimizing $E_h$ with respect to $\ta $ and $\tb $ yields the relations 
\begin{equation}
\label{v1}
\sin 2\ta +\sin (\ta -\tb )+\frac{1}{2}B'\sin \ta =0, 
\end{equation}
\begin{equation}
\label{v2}
-\eta \sin 2\tb -\sin (\ta -\tb )+\frac{1}{2}B'\sin \tb =0.
\end{equation}
In zero field, it is easy to show that $\tb =3\ta $ for all $\eta $. 
The equilibrium angles are plotted versus $\eta $ for several different values of
$B'$ ranging from 0 to 3 in the inset to Fig.2.  In the limit of large $\eta $ with $B'=0$, 
$\tb \rightarrow \pi /2$ and $\ta \rightarrow \pi /6$.  For nonzero field, $\tb $
still approaches $\pi /2$ but $\ta $ approaches an angle smaller than $\pi /6$.
In Fig.3(a), we plot the equilibrium angles versus $B'$ for several values of $\eta $.  

\begin{figure}
\includegraphics *[scale=0.7]{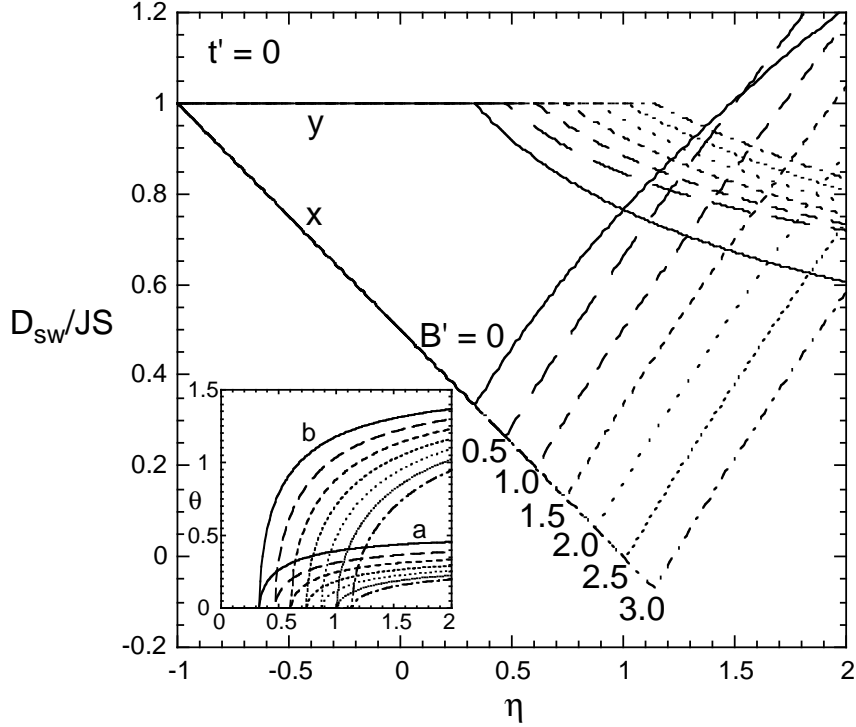}
\caption{
The SW stiffnesses versus $\eta $ for $t'=0$ and various values of the field $B'$.  In the
inset are plotted the equilibrium angles $\ta$ and $\tb $ versus $\eta $ for the same set of fields.
}
\end{figure}

After linearizing Eqs.(\ref{v1}) and (\ref{v2}), we find that the phase boundary 
between the CAF and FM phases satisfies the relation
\begin{equation}
\label{pb0}
B'-2\eta +4 -2\sqrt{(\eta +1)^2+1}=0,
\end{equation}
which was first obtained by Gabay {\it et al.} \cite{gab:89}.
While $\eta_c=1/3$ in zero field, $\eta_c$ increases with the field $B'$
as seen in the inset to Fig.2 and in Fig.3(a). 

\begin{figure}
\includegraphics *[scale=0.8]{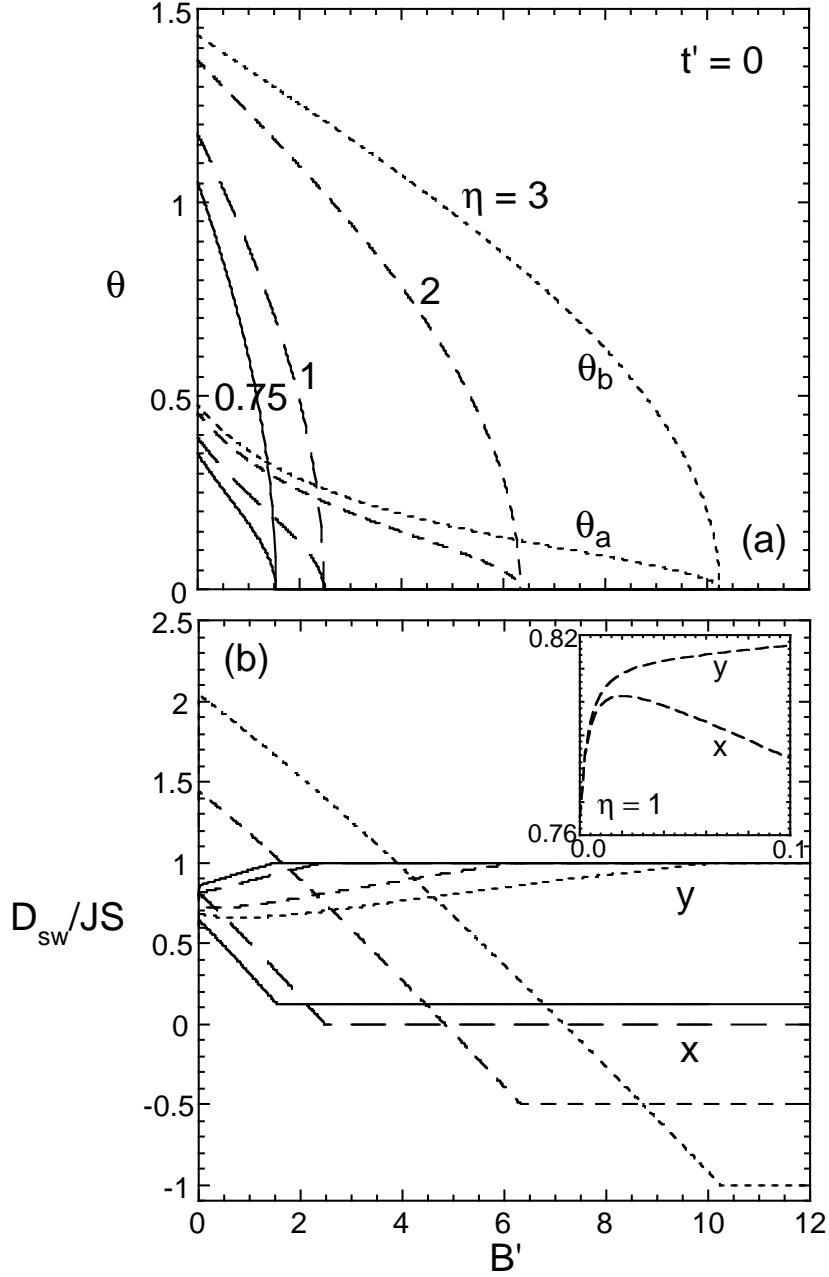}
\caption{
(a)  Equilibrium angles and (b) SW stiffnesses versus applied field $B'$ for several values 
of $\eta $ and $t'=0$.  Inset in (b) are the SW stiffnesses when $\eta =1$ for very small 
fields and $k_{\alpha }=0.015\pi $. 
}
\end{figure}

Expanded as $H_v =E_h+H_{v1}+H_{v2}+\ldots $ in powers of $1/\sqrt{S}$, the 
first-order term $H_{v1}$ vanishes provided that the angles $\ta $ and $\tb $ satisfy
Eqs.(\ref{v1}) and (\ref{v2}).  In terms of the Fourier-transformed spin operators
$a_{\vk }^{(r)}$ and $a_{\vk }^{(r) \dagger }$ on the $r=a$ or $b$ sublattice, 
the second-order term $H_{v2}$ can be written as 
\begin{equation}
\label{hv2} 
H_{v2}=JS\sum_{\vk, r,s}\biggl\{ a_{\vk }^{(r)\dagger }a_{\vk }^{(s)}A_{\vk }^{(r,s)}
+\Bigl( a_{-\vk }^{(r)}a_{\vk }^{(s)}+a_{-\vk }^{(r)\dagger }a_{\vk }^{(s)\dagger }\Bigr)
B_{\vk }^{(r,s)} \biggr\},
\end{equation}
\begin{equation}
A_{\vk }^{(a,a)}=2\cos 2\ta +2\cos (\ta -\tb ) -2\cos^2 \ta \cos k_x  
+B' \cos \ta , 
\end{equation}
\begin{equation}
A_{\vk }^{(a,b)}=A_{\vq }^{(b,a)}=-2 \cos^2 \bigl( (\ta -\tb )/2\bigr) \cos k_y , 
\end{equation}
\begin{equation}
A_{\vk }^{(b,b)}=2 \cos (\ta -\tb ) -2\eta \cos 2\tb +2\eta \cos^2 \tb \cos k_x
+B'\cos \tb , 
\end{equation}
\begin{equation}
B_{\vk }^{(a,a)}=  \sin^2 \ta \cos k_x , 
\end{equation}
\begin{equation}
B_{\vk }^{(a,b)}=B_{\vk }^{(b,a)}=  \sin^2 \bigl( (\ta -\tb )/2\bigr) \cos k_y , 
\end{equation}
\begin{equation}
B_{\vk }^{(b,b)}= - \eta \sin^2 \tb \cos k_x , 
\end{equation}
where the lattice constant is set to one.

The Hamiltonian of Eq.(\ref{hv2}) is easily diagonalized by applying the method of Walker and Walstedt 
\cite{wal:80}, which was originally developed for spin glasses.  The resulting spin excitation frequencies 
are given by
\begin{equation}
\label{sw}
\nu_{\vk }^{\pm } =\frac{JS}{\sqrt{2}}
\sqrt{ A_{\vk }^{(a,a) 2}+A_{\vk }^{(b,b) 2}+2\bigl(A_{\vk }^{(a,b) 2}-B_{\vk }^{(a,b) 2}\bigr)
-4\bigl(B_{\vk }^{(a,a) 2}+B_{\vk }^{(b,b) 2}\bigr) \pm R_{\vk }},
\end{equation}
\begin{eqnarray}
&R_{\vk }^2 = 4\Bigl\{ A_{\vk }^{(a,a) 2}+A_{\vk }^{(b,b) 2}-4\bigl(B_{\vk }^{(a,a) 2}+B_{\vk }^{(b,b) 2}\bigr)\Bigr\} 
\bigl(A_{\vk }^{(a,b) 2}-B_{\vk }^{(a,b) 2}\bigr)\nonumber \\ &
+\Bigl\{ A_{\vk }^{(a,a) 2}-A_{\vk }^{(b,b) 2}-4\bigl(B_{\vk }^{(a,a) 2}-B_{\vk }^{(b,b) 2}\bigr)\Bigr\}^2
+8\bigl(A_{\vk }^{(a,a)}A_{\vk }^{(b,b)}+4B_{\vk }^{(a,a)}B_{\vk }^{(b,b)}\bigr)\\ & 
\times \bigl(A_{\vk }^{(a,b) 2}+B_{\vk }^{(a,b) 2}\bigr)
-32A_{\vk }^{(a,b)}B_{\vk }^{(a,b)}\bigl(A_{\vk }^{(a,a)}B_{\vk }^{(b,b)}
+A_{\vk }^{(b,b)}B_{\vk }^{(a,a)}\bigr).\nonumber  
\end{eqnarray}
The SW frequency $\omega_{\vk }=\nu_{\vk }^{-}$ satisfies the 
condition $\omega_{\vk =0}=B$.  Our results for $\nu_{\vk }^{\pm }$ 
agree with the mode frequencies numerically evaluated by Saslow and Erwin \cite{sas:92}.
In the long-wavelength limit, the SW stiffnesses in the $x$ and $y$ directions
are obtained from the expression
$\lim_{\vk \rightarrow 0}\omega_{\vk }=B+\dsw^x k_x^2 +\dsw^y k_y^2$. 

In the FM phase, the SW frequency can be solved analytically:
\begin{equation}
\label{ok0}
\omega_{\vk }= B+JS\Bigl( 3-\eta +(\eta -1)\cos k_x \Bigr) 
-JS\sqrt{ (1+\eta )^2(1-\cos k_x)^2 +4\cos^2 k_y }.
\end{equation}
So for $k_x=0$, $\omega_{\vk }=B+2JS(1-\cos k_y)$ is independent of $\eta $.
The SW stiffnesses in the FM phase are given by the simple results $\dsw^x=(JS/2)(1-\eta )$ 
and $\dsw^y =JS $, independent of field.  The FM phase becomes unstable when $\omega_{\vQ }=0$, 
where $\vQ =(\pi ,0)$ is the AFM Bragg vector.  This yields the same condition for the 
CAF-FM phase boundary as Eq.(\ref{pb0}).

In the CAF phase, analytic results for the SW stiffnesses were found only when $B=0$:
\begin{equation}
\label{dswx0}
\dsw^x = JS \sqrt{2}\eta \sqrt{ 1-\ds\sqrt{\frac{\eta }{\eta +1}}},
\end{equation}
\begin{equation}
\label{dswy0}
\dsw^y = JS \sqrt{2}\sqrt{ 1-\ds\sqrt{\frac{\eta }{\eta +1}}}.
\end{equation}
When $\eta \rightarrow \infty $, $\dsw^x \rightarrow JS \sqrt{2\eta }$ and $\dsw^y\rightarrow JS\sqrt{2/\eta }$.
For $B > 0$, $\dsw^y$ tends to a nonzero limit but $\dsw^x $ still diverges as 
$\eta \rightarrow \infty $.  So the average SW stiffness 
always diverges when $\eta \rightarrow \infty $ regardless of the field.

As shown in Fig.2, $\dsw^x$ is a linearly decreasing function of $\eta $ in the FM 
phase below $\eta_c$.  In the CAF phase above $\eta_c$, $\dsw^x$ increases and 
$\dsw^y$ decreases with $\eta $.  Long-wavelength SW's become isotropic when 
$\dsw^x$ and $\dsw^y$ cross.  For $B=0$, $\dsw^x=\dsw^y$ when $\eta =1$.  
This crossing point moves to progressively larger values of $\eta $ with increasing field.   

The SW stiffnesses are plotted versus field in Fig.3(b) for four different values of $\eta $.  
In the FM phase above $B_c$, the stiffnesses are independent of field.  But in the CAF phase below $B_c$, 
the dependence on field is more complex.  For $\eta > 1$, $\dsw^x > \dsw^y$ at zero field and the stiffnesses cross 
as the field increases.  If the SW stiffness in the $\alpha $ direction is defined at fixed $k_{\alpha }$ by 
$\dsw^{\alpha }(k_{\alpha },B)=(\omega_{\vk } -B)/k_{\alpha }^2$, then $\dsw^{\alpha }(k_{\alpha },B)$ 
increases rapidly near the field $B^{\star }= \dsw^{\alpha }(k_{\alpha },0)k_{\alpha }^2$ as shown 
in the inset to Fig.3(b) for $k_{\alpha }=0.015\pi $.  As discussed in detail elsewhere \cite{fis:un},
this behavior is typical of any CAF with a quadratic SW dispersion.  Since the drop in $\dsw^x$ 
is then steeper than the rise in $\dsw^y$, $\dsw^{av}$ decreases with field for $B\gg B^{\star }$.

Surprisingly, Fig.3(b) indicates that the stiffness in the $x$ direction becomes negative
for sufficiently large $\eta $.  To understand this behavior, we have plotted the SW frequencies 
versus $\vk $ for $\eta =1$ and several fields in Fig.4.  Since $\omega_{\vQ }=0$ 
and $\omega_0=B$ in the CAF phase, the SW stiffness in the $x$ direction must decrease
as the field $B$ becomes comparable to $JS$.  When $\eta =1$, $\dsw^x =0$ for all fields above 
$B_c=2(\sqrt{5}-1)JS\approx 2.472JS$.  For $\eta > 1$ and $B > B_c$, 
$\dsw^x < 0$ in the FM phase.  Due to the reduced symmetry of the CAF phase, the first Brillouin 
zone extends from $-\pi $ to $\pi $ along $k_x$ but from $-\pi /2 $ to $\pi /2 $ along $k_y$
\cite{fbz}. 

\begin{figure}
\includegraphics *[scale=0.7]{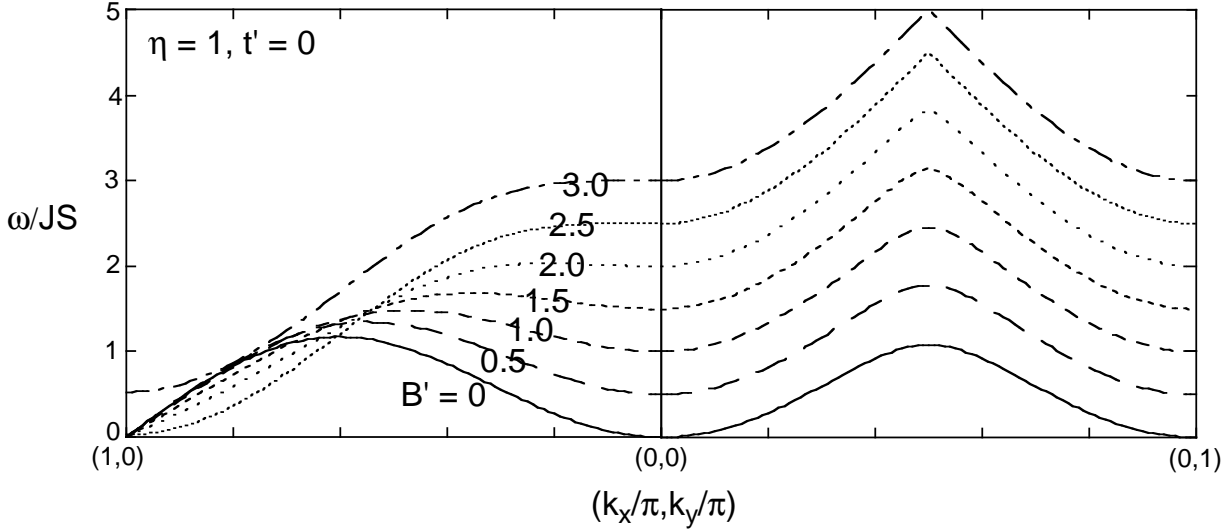}
\caption{
The SW frequencies for $p=0.66$, $\eta =1$, $t'=0$, and various fields $B'$.
}
\end{figure}

The SW frequencies are plotted versus $\vk $ for $B=0$ and various values of $\eta $ in Fig.5.
When $\eta =-1$, all of the Heisenberg interactions equal $J$ and the SW's are isotropic.  
For $\eta < \eta_c =1/3$ in the FM phase, the SW frequencies are independent of $\eta $
along $k_y$ but not along $k_x$, as implied by Eq.(\ref{ok0}).  With increasing $\eta $ in the CAF phase, 
the SW stiffness increases along the $x$ direction but decreases along the $y$ direction, as 
predicted by Eqs.(\ref{dswx0}) and (\ref{dswy0}) and shown in Fig.2.  Also notice that the SW 
velocity at $\vQ $ is an increasing function of $\eta $ in the CAF phase.

\begin{figure}
\includegraphics *[scale=0.7]{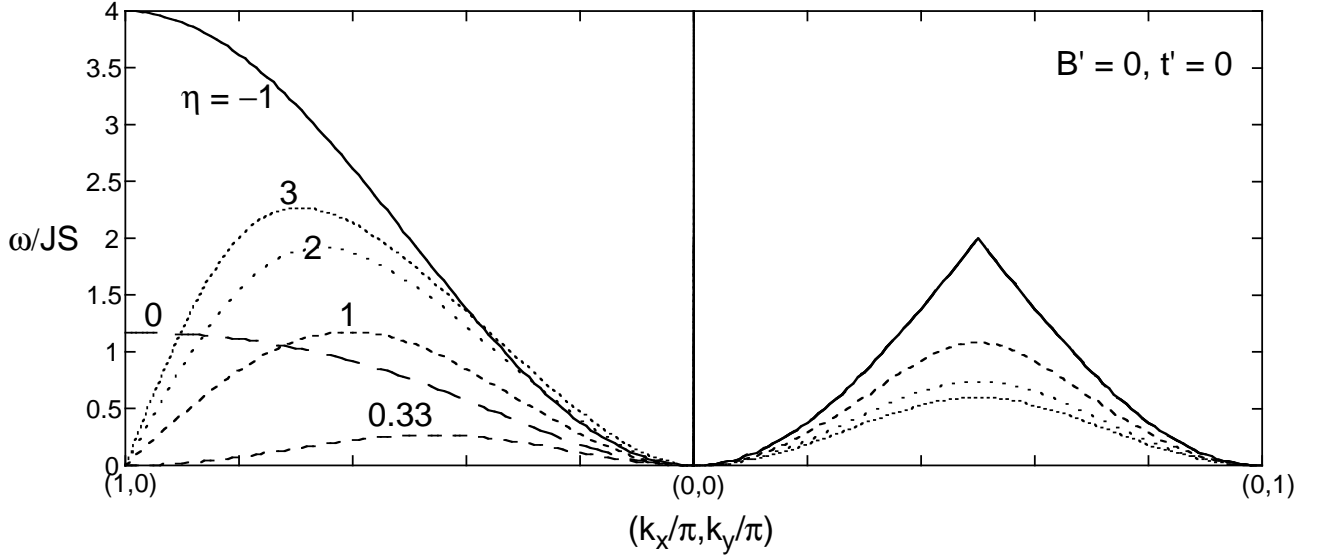}
\caption{
The SW frequencies for $B'=0$, $t'=0$, and several values of $\eta $.
}
\end{figure}

Both Figs.4 and 5 indicate that the SW velocity at $\vQ $ softens as the CAF
phase becomes unstable.  In the FM phase, $d\omega_{\vk }/dk_x$ 
vanishes at $\vQ $ for all $B > B_c$ or $\eta < \eta_c$.  The softening of the SW 
velocity at the CAF-FM phase boundary is a quite general result discussed by Rom\'an 
and Soto \cite{rom:00}. 

\section{Ground-state properties of the DEV model}

In this section, we discuss the ground-state properties of the DEV model, which
is constructed by placing electrons with density $p$ on the Villain lattice.  
An electron on site $i$ is coupled to the local moment at that site by Hund's coupling 
$-2\JH \vs_i \cdot \vS_i $ and is allowed to hop to neighboring sites with 
hopping energy $t$, as specified by the DEV Hamiltonian of Eq.(\ref{ham}).  

With the Fermion creation and destruction operators $\bc_{\vk \alpha }^{(r) \dagger }$ and 
$\bc_{\vk \alpha }^{(r)}$ defined in the reference frames of the local moments, the band 
Hamiltonian of the electrons can be written 
\begin{equation}
H_b = \sum_{\vk }\sum_{i,j }\sum_{\alpha ,\beta } \Bigl( H_{b0 \, \ab }^{ij}
+V_{\ab }^{ij}(\vk ) \Bigr) \bc_{\vk \alpha }^{(i) \dagger }\bc_{\vk \beta }^{(j)},
\end{equation}
where $\bc_{\vk \alpha }^{(i)}=\{ \bc_{\vk \alpha }^{(a)},\, \bc_{\vk+\vQ ,\alpha }^{(a)},\,
\bc_{\vk \alpha }^{(b)},\, \bc_{\vk+\vQ ,\alpha }^{(b)} \}$ defines the $ij$ subspace.
The sum over $\vk $ is restricted to the first Brillouin zone and $\alpha =\pm 1 $ corresponds 
to spin up or down in the local reference frames.  The zeroth-order band Hamiltonian includes 
just the Hund's coupling:  $H_{b0 \, \alpha \beta }^{ij} = -\JH S \delta_{ij} \sigma_{\ab }^z$.  
The potential $V_{\ab }^{ij}(\vk )$ is smaller by $t/\JH S$ and has matrix elements
\begin{equation}
V_{\ab }^{11}(\vk ) = -V_{\ab }^{22}(\vk )= -2t\cos k_x \cos \ta \dab , 
\end{equation}
\begin{equation}
V_{\ab }^{33}(\vk ) = -V_{\ab }^{44}(\vk )= -2t\cos k_x \cos \tb \dab , 
\end{equation}
\begin{equation}
V_{\ab }^{12}(\vk ) = -V_{\ab }^{21}(\vk )= -2ti \cos k_x \sin \ta \sigma^y_{\ab },
\end{equation}
\begin{equation}
V_{\ab }^{34}(\vk ) = -V_{\ab }^{43}(\vk )= -2ti \cos k_x \sin \tb \sigma^y_{\ab }.
\end{equation}
\begin{equation}
V_{\ab }^{13}(\vk ) = V_{\ab }^{24}(\vk ) = V_{\ab }^{31}(\vk )= V_{\ab }^{42}(\vk )=
-2t \cos k_y \cos ((\ta -\tb )/2) \dab , 
\end{equation}
\begin{equation}
V_{\ab }^{14}(\vk ) = V_{\ab }^{23}(\vk ) =  -V_{\ab }^{41}(\vk )= -V_{\ab }^{32}(\vk )= 
2ti \cos k_y \sin ((\ta -\tb )/2)\sigma^y_{\ab } ,
\end{equation}
Notice that the $\sigma^y_{\ab }$ terms couple the up and down spin states.
Like $E_h$ of Eq.(\ref{Eh}), $H_b $ is also of order $JS^2$.

The potential $V_{\ab }^{ij }(\vk )$ is treated within degenerate perturbation theory.  Second-order 
perturbation theory with corrections of order $t/\JH S$ will be required to obtain the SW frequencies in the 
next section.  But to order $(t/\JH S)^0$ or to first order in the potential, the spin up and down subspaces 
decouple and the band Hamiltonian $H_b$ is easily transformed into the diagonal form 
$H_b=\sum_{\vk ,\alpha, r}\epsilon_{\vk \alpha }^{(r)}d_{\vk \alpha }^{(r)\dagger } d_{\vk \alpha }^{(r)}$  
by the rotations $\bc_{\vk \alpha }^{(a)}=u_{\vk }^{(a)}d_{\vk \alpha }^{(a)} +u_{\vk }^{(b)}d_{\vk \alpha }^{(b)}$
and $\bc_{\vk \alpha }^{(b)}=u_{\vk }^{(b)}d_{\vk \alpha }^{(a)}-u_{\vk }^{(a)}d_{\vk \alpha }^{(b)}$
where $\epsilon_{\vk \alpha }^{(r)}=-\JH S \alpha +\te_{\vk }^{(r)}$, 
\begin{equation}
u_{\vk }^{(a) 2}=1-u_{\vk }^{(b) 2} = \ds\frac{1}{2} \biggl\{ 1 + 
\ds\frac{1}{\wk }\Bigl(\cos\theta_a -\cos\theta_b\Bigr)\cos k_x \biggr\} ,
\end{equation} 
\begin{equation}
\te_{\vk }^{(r)}=-t\cos k_x \Bigl(\cos \theta_a +\cos\theta_b \Bigr) \mp t \wk ,
\end{equation}
\begin{equation}
\wk =\sqrt{\Bigl(\cos \theta_a -\cos \theta_b\Bigr)^2\cos^2 k_x +4\cos^2 \bigl( (\theta_a-\theta_b)/2\bigr) \cos^2 k_y}. 
\end{equation}
The upper and lower signs in $\te_{\vk }^{(r)}$ refer to the $r=a$ and $b$ bands, respectively.  
With $H_b$ in diagonal form, $E_b=\langle H_b\rangle $ is easy to evaluate.

It is straightforward to minimize the zeroth-order energy $E_0=E_h+E_b$ with respect to the
angles $\ta $ and $\tb $ in the limit of large $\JH S$.  The relations that generalize 
Eqs.(\ref{v1}) and (\ref{v2}) are
\begin{equation}
\label{vv1} 
\sin 2\ta +\sin (\ta -\tb )+\frac{1}{2}B'\sin \ta +\frac{1}{NJS^2} \sum_{\vk }\Biggl\{
\frac{d\te_{\vk }^{(a)}}{d\ta }f(\te_{\vk }^{(a)}) +\frac{d\te_{\vk }^{(b)}}{d\ta }f(\te_{\vk }^{(b)})
\Biggr\}=0, 
\end{equation}
\begin{equation} 
\label{vv2} 
-\eta \sin 2\tb -\sin (\ta -\tb )+\frac{1}{2}B'\sin \tb +\frac{1}{NJS^2} \sum_{\vk }\Biggl\{
\frac{d\te_{\vk }^{(a)}}{d\tb }f(\te_{\vk }^{(a)}) +\frac{d\te_{\vk }^{(b)}}{d\tb }f(\te_{\vk }^{(b)})
\Biggr\}=0, 
\end{equation}
where
$f(\te )=\Theta (\tm -\te )$ is the Fermi function at $T=0$ and $\tm = \mu -\JH S \, \sgn (p-1)$ 
is the shifted chemical potential.  For a fixed $\eta $ and $B'$, the equilibrium angles decrease 
with increasing $t'$.  The phase boundary between the CAF and FM phases is now given by
the condition
\begin{equation}
\label{pb}
B'-2\eta +4 +3\eke /4JS^2-2\sqrt{ (1+\eta )^2+1 +\eke /4JS^2 +(\eke /8JS^2)^2 }=0,
\end{equation}
where $\eke =-\bigl(\langle \te^{(a)}_{\vk }\rangle +\langle \te^{(b)}_{\vk }\rangle \bigr)/2 \ge 0$ is 
the average kinetic energy of the electrons in the FM phase.  This reduces
to Eq.(\ref{pb0}) when $\eke =0$.  For $p=0.66$, $\eta =2$, and $B'=0$, the dependence of the equilibrium 
angles on $t'$ is plotted in Fig.6(a).  Also shown is the average spin $M=S(\cos\theta_a +\cos\theta_b)/2$ 
of the local moments.  When $t'=10$ and $B'=0$, the equilibrium angles and $M/S$ are plotted versus 
$\eta $ in Fig.7(a).  Their dependence on field $B'$ is plotted in Fig.8(a) for $t'=3$ and $\eta =2$. 

\begin{figure}
\includegraphics *[scale=0.8]{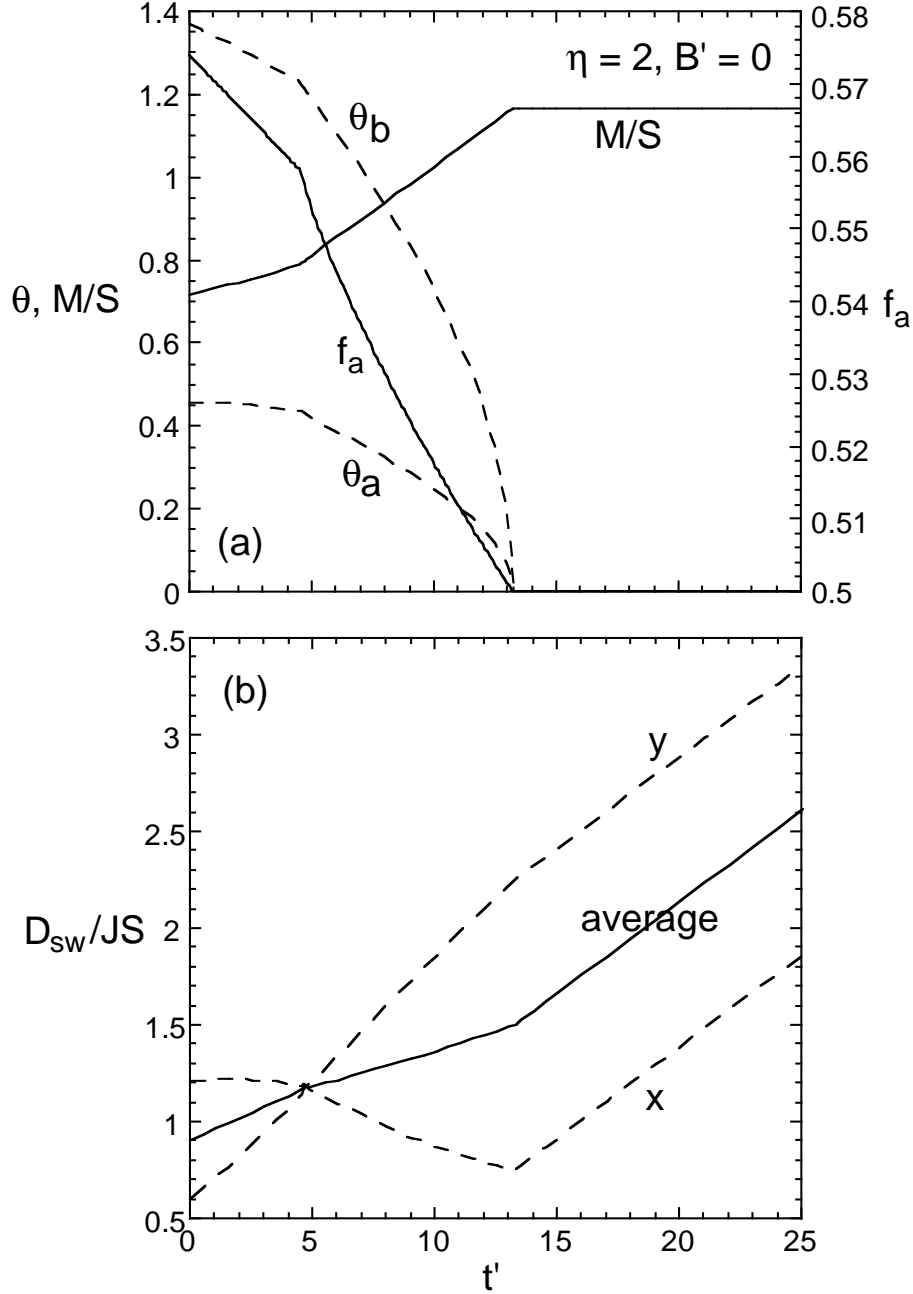}
\caption{
(a) The angles $\theta_a$ and $\theta_b$ of the local moments, the total local magnetization
$M/S$, the fraction $f_a$ of the electrons on the $a$ sites, and (b) the spin-wave stiffnesses
(both in the $x$ and $y$ directions and their average) for $p=0.66$, $\eta =2 $, and $B'=0$ versus $t'$.  
}
\end{figure}

\begin{figure}
\includegraphics *[scale=0.8]{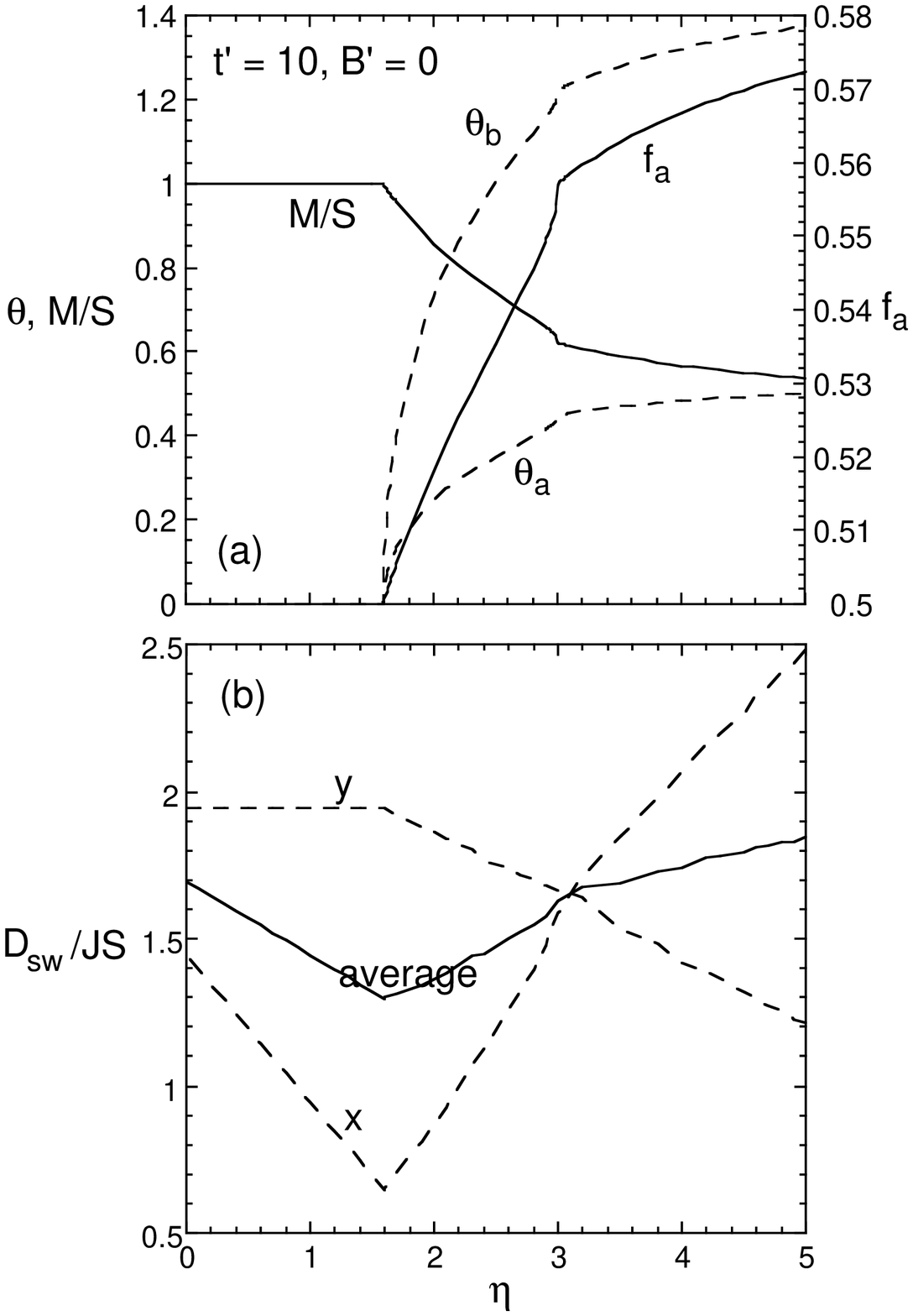}
\caption{
(a) The angles $\theta_a$ and $\theta_b$ of the local moments, the total local magnetization
$M/S$, the fraction $f_a$ of the electrons on the $a$ sites, and (b) the spin-wave stiffnesses
for $p=0.66$, $t'=10$, and $B'=0$ versus $\eta $.  
}
\end{figure}
 
\begin{figure}
\includegraphics *[scale=0.8]{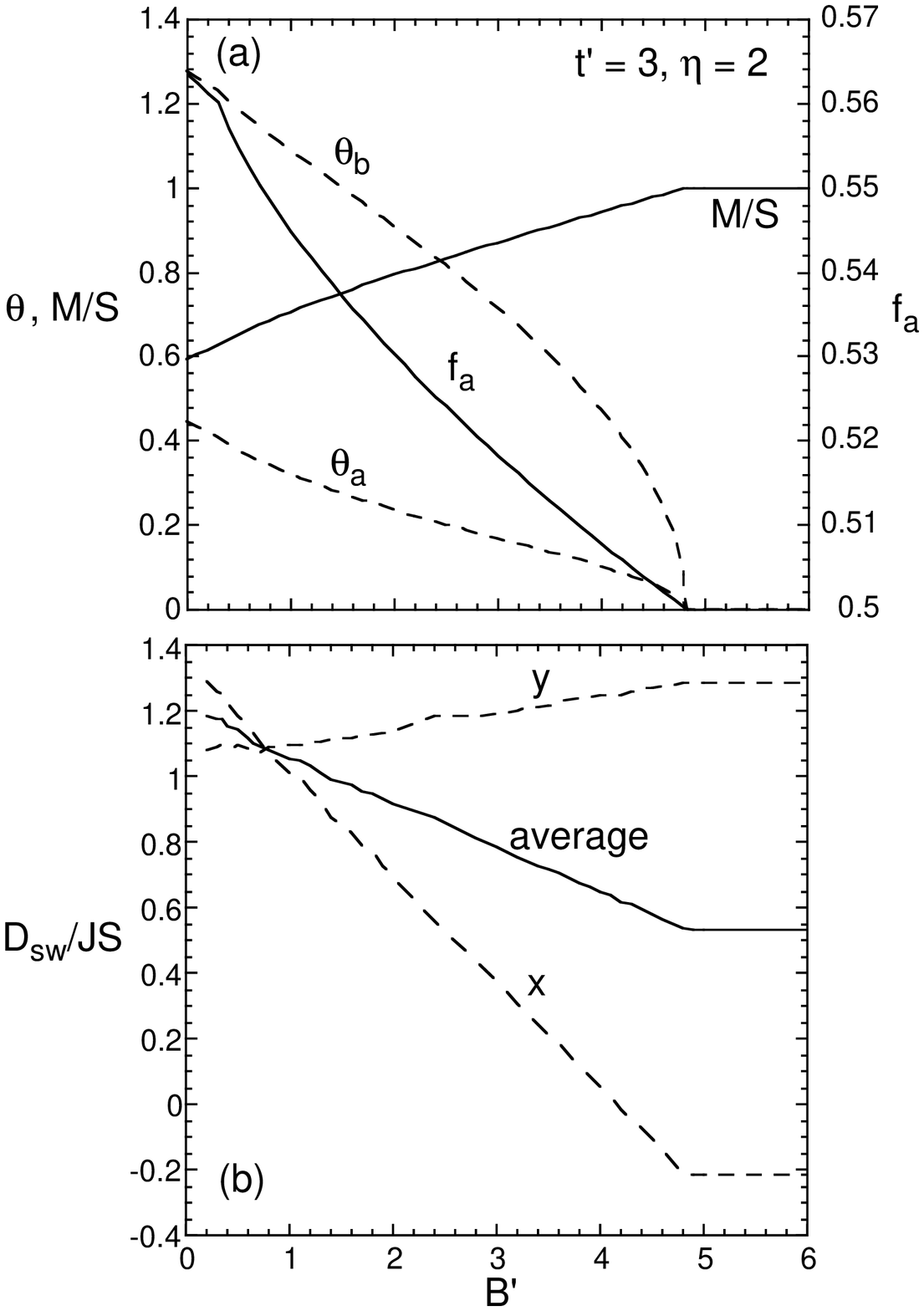}
\caption{
(a) The angles $\theta_a$ and $\theta_b$ of the local moments, the total local magnetization
$M/S$, the fraction $f_a$ of the electrons on the $a$ sites, and (b) the spin-wave stiffnesses
for $p=0.66$, $t'=3$, and $\eta =2$ versus $B'$.  
}
\end{figure}

Surprisingly, the electronic occupation of the $a$ and $b$ sites on the Villain
lattice are different.  For $p < 1$, the occupancies of the $a$ and $b$ sublattices are given by
\begin{equation}
n_a=\ds\frac{2}{N}\sum_{\vk } \Bigl\{ u_{\vk }^{(a) 2} f(\te_{\vk }^{(a)})+u_{\vk }^{(b) 2} 
f(\te_{\vk }^{(b)})\Bigr\},
\end{equation}
\begin{equation}
n_b=\ds\frac{2}{N}\sum_{\vk } \Bigl\{ u_{\vk }^{(b) 2} f(\te_{\vk }^{(a)})+u_{\vk }^{(b) 2} 
f(\te_{\vk }^{(a)})\Bigr\},
\end{equation}
From the relation $u_{\vk }^{(a) 2}+u_{\vk }^{(b) 2}=1$, it follows that $(n_a+n_b)/2=p$ 
is just the average number of electrons per site.  

In the CAF phase, electrons prefer to sit on the $a$ sites of the Villain lattice.
The fraction $f_a=n_a/2p \ge 1/2$ of such electrons is plotted in Figs.(6-8).
For $\eta =2 $ and $B=0$ in Fig.6(a), $f_a$ has a maximum of 0.574 as $t'\rightarrow 0$ 
and approaches 1/2 as $t'\rightarrow t'_c\approx 13.2$.  Similar behavior is found in 
Figs.7(a) and 8(a), where $f_a$ is shown to be an increasing function of $\eta $ and a decreasing function of
$B'$.  This behavior is easy to understand:  the largest angles between neighboring spins 
are along the $x$ axis between $b$ sites with angles differing by $2\theta_b $.  When an 
electron hops onto a $b$ site, it cannot easily hop to other $b$ sites and so quickly moves 
onto a neighboring $a$ site, where it can readily travel between other $a$ sites with angular 
difference $2\theta_a \ll 2\theta_b$.  Hence, the non-collinearity of the local moments quite 
naturally produces a charge-density wave (CDW) with a substantial amplitude.  As $\eta \rightarrow
\infty $, $\tb $ approaches $\pi /2$ and $f_a$ approaches $0.593$.  So even 
when the electrons are unable to hop between sites on the $b$ sublattice (since the angles on 
neighboring $b$ sites differ by $\pi $), roughly 40\% of the electrons can still be found on $b$
sites at any one time.  Because neither the CAF nor FM densities-of-states contain a gap, both
phases are metallic within the DEV model for $t' > 0$. 

Due to short-range orbital ordering, a CDW with the same period as the one predicted here has in 
fact been observed in the AFM regions of Pr$_{0.7}$Ca$_{0.3}$MnO$_3$ \cite{oki:99,asa:02}.  However, 
the observed charge ordering is perfect:  all of the Mn$^{3+}$ ions lie on one sublattice and
all of the Mn$^{4+}$ ions lie on the other.  Such perfect charge ordering is never achieved 
within the DEV model.    

Another surprising result is that phase separation occurs within a narrow range of $t'$.  
In a plot of filling $p$ versus chemical potential $\mu $, phase separation appears 
as a jump $\Delta p $ in $p(\mu )$.  For the parameters $\eta =3$ and $B=0$ in Fig.9, 
$\Delta p$ reaches a maximum of about 0.0028 when $t'\approx 10.0$ and shrinks as $t'$ increases.  
If $p$ is fixed at 0.66, then phase separation occurs within a very narrow range of $t'$ 
between about 9.98 and 10.02.  For fixed $p$, phase separation appears as jumps
in the equilibrium angles $\theta_r$ and in the electron fraction $f_a$, as seen in Figs.6(a) 
and 7(a) for zero field.  

\begin{figure}
\includegraphics *[scale=0.8]{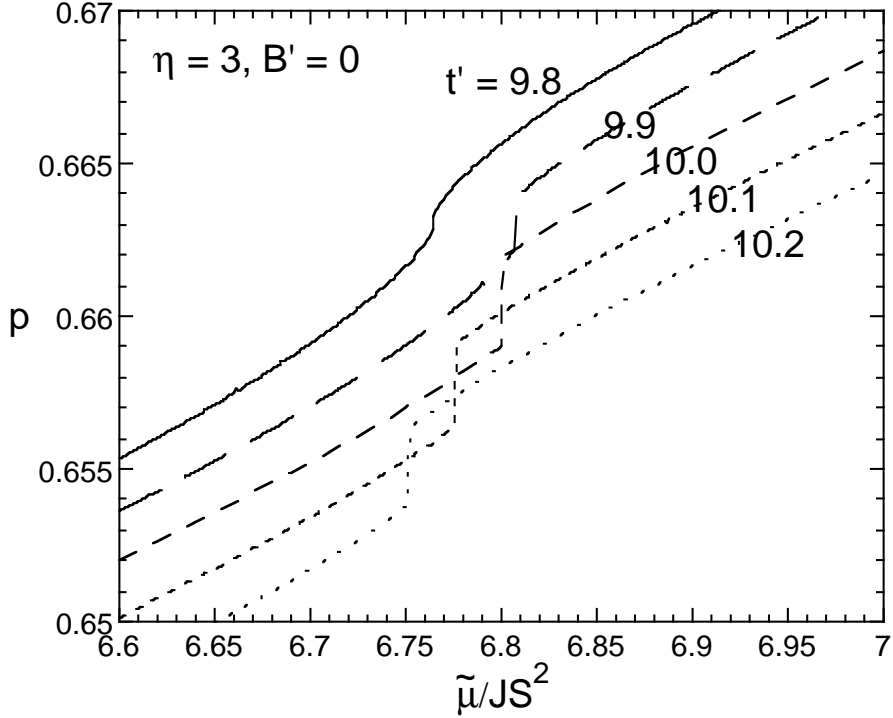}
\caption{
The filling $p$ versus chemical potential $\tilde \mu $ for various values of $t'$, 
$\eta =3$, and $B'=0$.  Phase separation appears as discontinuities in $p$.
}
\end{figure}

Like the Pomeranchuk instability \cite{hal:00,val:01} in the two-dimensional Hubbard model, 
the phase instability in the DEV model occurs close to but slightly above the Van Hove filling and 
is marked by a change in Fermi surface (FS) topology from closed to open.  In the usual 
Pomeranchuk instability, however, the change in FS topology spontaneously breaks the 
square symmetry of the lattice.  Square symmmetry is already broken in the DEV model
by the Villain arrangement of the Heisenberg interactions.  The FS of the DEV model is sketched in 
Fig.10 for $\eta =3$, $B=0$, $p=0.66$, and for values of $t'$ on either side of the phase-separated 
range.  For $t'$ just above 10.0, the extra electrons in the neck of the $a$ FS around $\vk =0$ are 
offset by the holes in the $b$ FS around $\vk =(\pi ,\pi/2)$.  Although phase separation occurs 
for any $t'$ around some value of the filling, it becomes significant only if the $b$ FS is already present 
when the necks in the $a$ FS develop.  For the parameters in Fig.9, this requires that $t' \ge 9.9$.  

\begin{figure}
\includegraphics *[scale=0.6]{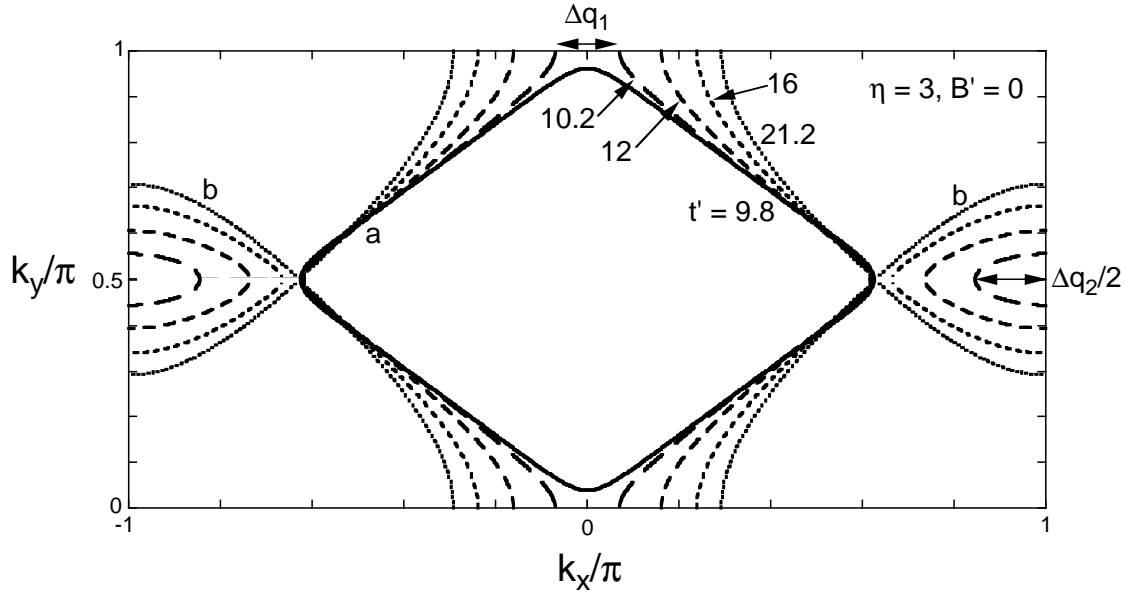}
\caption{
The FS is plotted for $\eta =3$, $B'=0$, $p=0.66$, and $t' =9.8$, 10.2, 12, 16, or 21.2.  
For $t' > 10$, the $a$ FS develops necks around $\vk =0$. 
}
\end{figure}

A magnetic field very quickly narrows and then eliminates the region of phase separation.  
For $t'=3$ and $\eta =2$, phase separation does not occur with increasing
field.  The small kinks in the equilibrium angles and electron fraction seen in Fig.8(a)
correspond to points of Van Hove filling (where the necks in the $a$ FS first appear)
but not to a phase-separated region. 

\section{Spin dynamics of the DEV model}

In this section, we evaluate the SW frequencies for the DEV model by following the general formalism 
developed by Golosov \cite{gol:00}.  While introducing some new physics, the presence of non-collinear 
spins also complicates things a bit. 

When $\JH S/t$ is finite, the relationships given above for $\bc_{\vk \alpha }^{(r)}$ in 
terms of $d_{\vk \alpha }^{(s)}$ must contain admixtures of the opposite-spin terms 
$(t/\JH S)d_{\vk +\vQ, -\alpha }^{(s)}$.  As sketched in Fig.1(c), this implies that the equilibrium 
angles $\psi_r$ of the electrons are smaller than the angles $\theta_r$ of the local moments
because the electrons try to align their spins as much as possible.  For large Hund's coupling, 
$\theta_r-\psi_r\propto t/\JH S$ and the electrons always exert a small torque on the local moments.  
In the local reference frame of site $i$ on the $r$ sublattice, $\langle \bas_{ix} \rangle 
=(n_r/2) \sin (\theta_r -\psi_r )$ so these new terms produce a correction to the Hund's 
coupling $-\JH \vS_i \cdot \vs_i $ that survives in the $\JH S \rightarrow \infty $ limit.  
Indeed, these torque terms are required to obtain sensible results for the SW frequencies.

To second order in perturbation theory, the new relationships for the Fermion operators
are given by
\begin{eqnarray}
\bc_{\vk \alpha }^{(a)}&=u_{\vk }^{(a)}d_{\vk \alpha }^{(a)} +u_{\vk }^{(b)}d_{\vk \alpha }^{(b)}
+\ds\frac{t}{\JH S}\Biggl\{ \Bigl( u_{\vk }^{(a)} \cos k_y \sin \bigl( (\ta -\tb )/2 \bigr) 
-u_{\vk }^{(b)} \cos k_x \sin \ta \Bigr) d_{\vk +\vQ ,-\alpha }^{(a)} 
\nonumber \\ &
-\Bigl( u_{\vk }^{(a)} \cos k_x \sin \ta +u_{\vk }^{(b)} \cos k_y \sin \bigl( (\ta -\tb )/2 \bigr)
\Bigr) d_{\vk +\vQ ,-\alpha }^{(b)}\Biggr\} ,
\end{eqnarray}
\begin{eqnarray}
\bc_{\vk \alpha }^{(b)}&=u_{\vk }^{(b)}d_{\vk \alpha }^{(a)}-u_{\vk }^{(a)}d_{\vk \alpha }^{(b)}
+\ds\frac{t}{\JH S}\Biggl\{ -\Bigl( u_{\vk }^{(a)} \cos k_x \sin \tb +u_{\vk }^{(b)} \cos k_y 
\sin \bigl( (\ta -\tb )/2 \bigr) \Bigr) d_{\vk +\vQ ,-\alpha }^{(a)} 
\nonumber \\ &
+\Bigl( -u_{\vk }^{(a)} \cos k_y \sin \bigl( (\ta -\tb )/2 \bigr) +u_{\vk }^{(b)} \cos k_x \sin
\tb \Bigr)  d_{\vk +\vQ ,-\alpha }^{(b)} \Biggr\} . 
\end{eqnarray}
On the $r$ sublattice, these equations imply that $\theta_r-\psi_r$ is related to the band and 
harmonic energies by 
\begin{equation}
\frac{n_r}{2}\sin (\theta_r -\psi_r ) =\frac{1}{\JH S}\frac{d}{d\theta_r}\frac{E_b}{N}
=-\frac{1}{\JH S}\frac{d}{d\theta_r}\frac{E_h}{N},
\end{equation}
which can also be obtained by minimizing the classical energy ${\cal E}(\ta ,\tb )
=-\JH S\sum_i n_i \cos (\theta_i -\psi_i )+E_h(\ta ,\tb)$ with respect to $\ta $ or $\tb $. 
Hence, the torque exerted by the electrons on the local moments opposes the tendency of the
local moments to return to the angles that minimize $E_h$. 
  
In terms of the Fermion operators $d_{\vk \alpha }^{(r)}$ and $d_{\vk \alpha }^{(r)\dagger }$
that diagonalize $H_b$, the full Hamiltonian can be expanded in a power series in $1/\sqrt{S}$ as 
$H=H_0+H_1+H_2+\cdots $.  The first-order term may be written 
\begin{eqnarray}
\label{h1}
&H_1= -2\JH\ds\sqrt{\frac{S}{N}}\sum_{\vk ,\vq, r, s}d_{\vk +\vq ,\da }^{(r) \dagger }
d_{\vk ,\ua }^{(s)} \biggl\{ u_{\vk +\vq }^{(r)}u_{\vk }^{(s)} a_{\vq }^{(a)}
+ v_{\vk +\vq }^{(r)}v_{\vk }^{(s)} a_{\vq }^{(b)} \biggr\} \nonumber \\ &
-\ds\frac{2t}{\sqrt{SN}}\sum_{\vk ,\vq ,r,s}\Biggl\{ d_{\vk +\vq ,\ua }^{(r) \dagger }d_{\vk +\vQ ,\ua }^{(s)}
\Bigl( u_{\vk +\vq }^{(r)}x_{\vk }^{(s)}a_{\vq }^{(a)}+ v_{\vk +\vq }^{(r)}y_{\vk }^{(s)}a_{\vq }^{(b)}
\Bigr) \nonumber \\ &
+ d_{\vk +\vq +\vQ ,\da }^{(r) \dagger }d_{\vk \da }^{(s)}
\Bigl( u_{\vk }^{(s)}x_{\vk +\vq }^{(r)}a_{\vq }^{(a)}+ v_{\vk }^{(s)}y_{\vk +\vq }^{(r)}a_{\vq }^{(b)}
\Bigr)\Biggr\} +h.c., 
\end{eqnarray}
where 
\begin{equation}
x_{\vk }^{(a)}=u_{\vk }^{(a)}\cos k_y \sin \bigl( (\theta_a-\theta_b)/2\bigr) -u_{\vk }^{(b)} \cos k_x \sin\theta_a,
\end{equation}
\begin{equation}
y_{\vk }^{(a)}=-u_{\vk }^{(b)}\cos k_y \sin \bigl( (\theta_a-\theta_b)/2\bigr) -u_{\vk }^{(a)} \cos k_x \sin\theta_b,
\end{equation}
and $x_{\vk }^{(b)}=-x_{\vk +\vQ}^{(a)}$, $y_{\vk }^{(b)}=y_{\vk +\vQ}^{(a)}$, $v_{\vk }^{(a)}=u_{\vk }^{(b)}$,
and $v_{\vk }^{(b)}=-u_{\vk }^{(a)}$.   The second term in Eq.(\ref{h1}) is produced by the torque exerted
by the electrons on the local moments.  Notice that $H_1$ is linear in the boson operators.  The 
expectation value of the Fermion factor multiplying the boson operators in Eq.(\ref{h1}) vanishes provided 
that $\theta_r$ satisfy Eqs.(\ref{vv1}) and (\ref{vv2}).  The second-order Hamiltonian can be written as
\begin{equation}
H_2= \frac{2\JH }{N} \sum_{\vk ,\vq_1 ,\vq_2 ,r ,s}\sum_{\alpha ,\beta } \sigma^z_{\ab }
d_{\vk -\vq_1 ,\alpha }^{(r) \dagger } d_{\vk -\vq_2 ,\beta }^{(s) }
\Biggl\{ a_{\vq_1 }^{(a)\dagger }a_{\vq_2}^{(a)} u_{\vk -\vq_1 }^{(r)} u_{\vk -\vq_2 }^{(s)} 
+ a_{\vq_1 }^{(b)\dagger }a_{\vq_2}^{(b)} v_{\vk -\vq_1 }^{(r)} v_{\vk -\vq_2 }^{(s)}\Biggr\}. 
\end{equation}
Since $\JH /JS$ and $t/JS$ are of order $S^0$, $H_1/JS^2$ is of order $1/\sqrt{S}$ and $H_2/JS^2$ is of order $1/S$.

To eliminate the first-order term in $H$ and to express the Hamiltonian in terms of the true SW operators 
for the total spin $\vS_{i, {\rm tot}}=\vS_i+\vs_i$, we perform the unitary transformation \cite{gol:00} 
$H'=e^{-U}He^U$ where $U$ is constructed to satisfy $[U,H_0]=H_1$.  This transformation produces a 
modified second-order term $H_2'=H_2+[U,H_1]/2$.  The anti-Hermitian operator $U$ that fulfills these 
requirements is
\begin{eqnarray}
&U=-2\JH \sqrt{\ds\frac{S}{N}}\ds\sum_{\vk ,\vq ,r ,s}\Biggl\{ d_{\vk +\vq ,\da }^{(r) \dagger }d_{\vk \ua }^{(s)} 
\Bigl( a_{\vq }^{(a)} u_{\vk +\vq }^{(r)}u_{\vk }^{(s)} +a_{\vq }^{(b)} v_{\vk +\vq }^{(r)} v_{\vk }^{(s)} \Bigr)
\ds\frac{1}{\te_{\vk }^{(s)}-\te_{\vk +\vq }^{(r)}-2\JH S}\Biggr\}
\nonumber \\ &
-\ds\frac{2t}{\sqrt{SN}}\ds\sum_{\vk ,\vq ,r ,s}\Biggl\{ d_{\vk +\vq ,\ua }^{(r) \dagger }d_{\vk +\vQ ,\ua }^{(s)}
\Bigl( a_{\vq }^{(a)} u_{\vk +\vq }^{(r)}x_{\vk }^{(s)} +a_{\vq }^{(b)} v_{\vk +\vq }^{(r)}y_{\vk }^{(s)}\Bigr)
\nonumber \\ &
+ d_{\vk +\vq ,\da }^{(r) \dagger }d_{\vk +\vQ ,\da }^{(s)}
\Bigl( a_{\vq }^{(a)} u_{\vk +\vQ }^{(s)}x_{\vk +\vq +\vQ }^{(r)} +a_{\vq }^{(b)} v_{\vk +\vQ }^{(s)}y_{\vk +\vq +\vQ }^{(r)}\Bigr)
\Biggr\}
\ds\frac{1}{\te_{\vk +\vQ }^{(s)}-\te_{\vk +\vq }^{(r)}} - h.c.,
\end{eqnarray}
which is clearly of order $(JS^2)/\sqrt{S}$.

Performing the unitary transformation and taking the limit $\JH S\rightarrow \infty $ produces a rather
complicated expression for $H_2'$.  This expression can be considerably simplified by
tracing over the Fermion degrees of freedom, thereby replacing the combination of Fermi operators 
$d_{\vk \uparrow }^{(r) \dagger }d_{\vk' \uparrow }^{(s)}$ by its expectation value 
$\delta_{rs}\delta_{\vk ,\vk'} f(\te_{\vk }^{(r)})$.  Then $H_2'$ can be written 
as an effective Hamiltonian for the SW operators $a_{\vq }^{(r)}$ and $a_{\vq }^{(r)\dagger }$
that takes the same form as the harmonic Hamiltonian of Eq.(\ref{hv2}) but with revised coefficients
$\tilde A_{\vk }^{(r,s)}=A_{\vk }^{(r,s)}+C_{\vk }^{(r,s)}+D_{\vk }^{(r,s)}$ and 
$\tilde B_{\vk }^{(r,s)}=B_{\vk }^{(r,s)}+E_{\vk }^{(r,s)}$.  The new contributions
to these coefficients are
\begin{eqnarray}
C_{\vk }^{(r,s)}=& -\ds\frac{1}{N}\sum_{\vq ,l}f(\te_{\vq }^{(l)})\Biggl\{ 
u_{\vq }^{(l) 2} \Bigl( 2t' \cos \theta_a \cos (k_x +q_x) +\te_{\vq }^{(l)}\Bigr) \delta_{r,a}\delta_{s,a}\nonumber \\ & 
+v_{\vq }^{(l) 2} \Bigl( 2t' \cos \theta_b \cos (k_x +q_x) +\te_{\vq }^{(l)}\Bigr) \delta_{r,b}\delta_{s,b}\nonumber \\ & 
+2t'\cos \bigl( (\theta_a -\theta_b)/2\bigr)
\cos (k_y +q_y) u_{\vq }^{(l)} v_{\vq }^{(l)} (\delta_{r,a}\delta_{s,b}+\delta_{r,b}\delta_{s,a})\Biggr\},
\end{eqnarray}
\begin{eqnarray}
D_{\vk }^{(r,s)}=&\ds\frac{4{t'}^2JS^2}{N}\sum_{\vq ,l,m}\Bigl( 
u_{\vk +\vq }^{(l)}x_{\vq }^{(m)}\delta_{r,a}
+v_{\vk +\vq }^{(l)}y_{\vq }^{(m)}\delta_{r,b}
\Bigr) \nonumber \\ &
\Bigl( 
u_{\vk +\vq }^{(l)}x_{\vq }^{(m)}\delta_{s,a} 
+v_{\vk +\vq }^{(l)}y_{\vq }^{(m)}\delta_{s,b} 
\Bigr)
\ds\frac{ f(\te_{\vq +\vQ}^{(m)})-f(\te_{\vk +\vq}^{(l)})}{\te_{\vq +\vQ}^{(m)}-\te_{\vk +\vq}^{(l)}},
\end{eqnarray}
\begin{eqnarray}
E_{\vk }^{(r,s)}=&\ds\frac{2{t'}^2JS^2}{N}\sum_{\vq ,l,m}\Bigl( 
u_{\vq +\vQ }^{(m)}x_{\vk +\vq +\vQ }^{(l)}\delta_{r,a}
+v_{\vq +\vQ }^{(m)}y_{\vk +\vq +\vQ }^{(l)}\delta_{r,b}
\Bigr)\nonumber \\ &
\Bigl( 
u_{\vk +\vq }^{(l)}x_{\vq }^{(m)}\delta_{s,a} 
+v_{\vk +\vq }^{(l)}y_{\vq }^{(m)}\delta_{s,b}
\Bigr)
\ds\frac{ f(\te_{\vq +\vQ}^{(m)})-f(\te_{\vk +\vq}^{(l)})}{\te_{\vq +\vQ}^{(m)}-\te_{\vk +\vq}^{(l)}}.
\end{eqnarray}
Spin excitation frequencies $\nu_{\vq }^{\pm }$ are given by Eq.(\ref{sw}) with $\tilde A_{\vk}^{(r,s)}$
and $\tilde B_{\vk }^{(r,s)}$ replacing the harmonic coefficients.  Note that the torque terms in $H_1$ produce 
$D_{\vq }^{(r,s)}$ and $E_{\vq }^{(r,s)}$, both of which vanish in the FM phase.  In the CAF phase, 
these terms are required to preserve rotational symmetry and the relations $\omega_{\vQ }=0$ and $\omega_0=B$. 

An analytic expression for the SW frequency is possible in the FM phase:
\begin{eqnarray}
\label{ok}
\omega_{\vk }= &B+JS\Bigl( 3-\eta +(\eta -1)\cos k_x \Bigr) +\eke (2-\cos k_x)/4S
\nonumber \\ &
-JS\sqrt{ (1+\eta )^2(1-\cos k_x)^2 +(2+\eke /4JS^2)^2\cos^2 k_y },
\end{eqnarray}
which generalizes Eq.(\ref{ok0}) for the Villain model.
The FM phase becomes unstable when $\omega_{\vQ }=0$, which yields the same condition for
the CAF-FM phase boundary as Eq.(\ref{pb}).  In the FM phase, the SW stiffnesses are the 
sum of DE and Heisenberg contributions:  $\dsw^x=\eke /8S+JS(1-\eta )/2$ and $\dsw^y=\eke /8S+JS$,
both independent of field.  For $J=0$, these results agree with the SW frequencies of the DE model 
first obtained by Furukawa \cite{fur:95}.  When $\eta =-1$, the SW's are isotropic because all 
nearest-neighbor interactions equal $J$. 

In the CAF phase, the SW frequency and stiffnesses are solved numerically by integrating over the first 
Brillouin zone to obtain $C_{\vk }^{(r,s)}$, $D_{\vk }^{(r,s)}$, and $E_{\vk }^{(r,s)}$.  
Results for the SW frequency are plotted in Fig.11 for $\eta =3$, $B'=0$, and 
several values of $t'$.  For all $t' \le t'_c\approx 21.2$, the SW frequency vanishes
at both the FM and AFM Bragg vectors.  Above the phase separation region around $t' \approx 10.0$ but
below $t'_c$, $\omega_{\vk }$ develops kinks that correspond to transitions across the neck of the $a$ FS
($\Delta \vq_1 = \vQ -\vk_1 \approx 0.14 \pi \vx $ for $t'=10.2$) and across the length of the $b$ FS ($\Delta \vq_2
= \vQ - \vk_2 \approx 0.33\pi \vx $ for $t'=10.2$).  These transitions are sketched in Fig.10.  Since
the nesting across $\Delta \vq_2$ is much weaker than across $\Delta \vq_1$, the kink in the SW frequency
at $\vk_2 $ is much weaker than the one at $\vk_1$.  As $t' \rightarrow 10.0$, $\Delta \vq_i \rightarrow 0$ 
so the kinks at $\vk_1$ and $\vk_2$ merge with $\vQ $.  
Whereas $\omega_{\vk }$ is a non-monotonic function of $t'$ along the $x$ 
direction, it is a monotonically increasing function of $t'$ along the $y$ direction.  

\begin{figure}
\includegraphics *[scale=0.65]{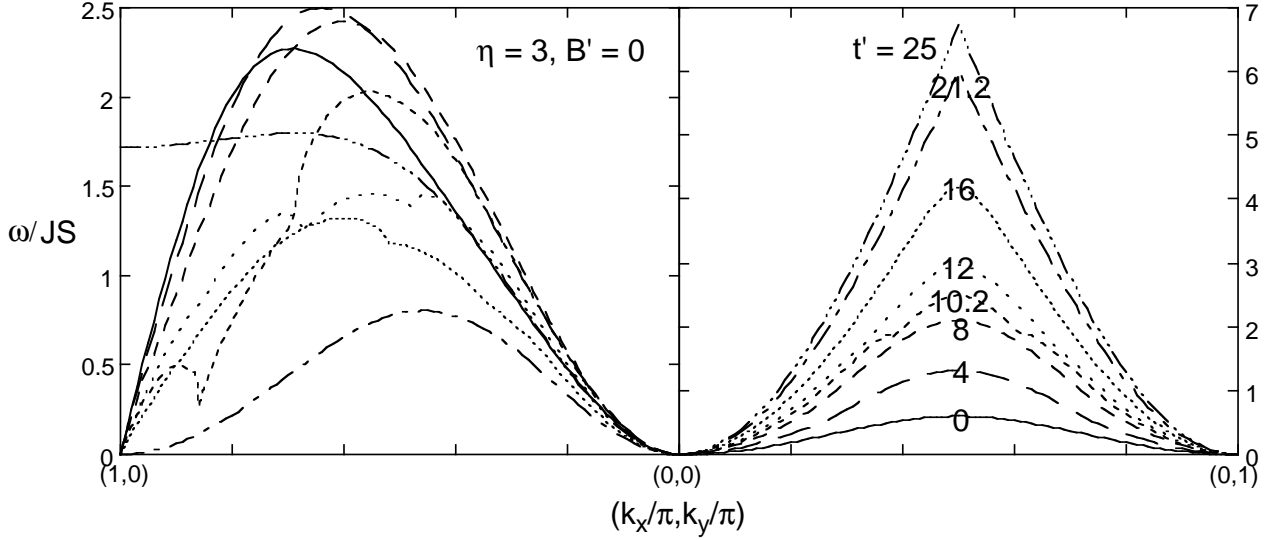}
\caption{
The SW frequencies for $p=0.66$, $\eta=3$, $B'=0$, and various values of $t'$.  
Different frequency scales are used on either side of $\vk =0$.
}
\end{figure}

As plotted in Fig.6(b) for $\eta =2$, the SW stiffness in the $x$ direction reaches a 
minimum at $t'_c\approx 13.2$, above which both $\dsw^x$ and $\dsw^y$ are linearly increasing 
functions of $t'$.  The stiffnesses in the $x$ and $y$ directions cross in the region of phase separation,
where the SW's become isotropic.  Notice that $\dsw^{av}$ increases by about 65\% 
as $t'$ increases from zero to $t'_c$.  For $\eta =3$, $\dsw^{av}$ doubles in this interval \cite{fis:04}

With increasing $\eta $, the SW stiffnesses plotted in Fig.7(b) for $t'=10$ and $B=0$ 
again cross in the region of phase separation around $\eta = 3.0$.  For larger $\eta $, 
$\dsw^x > \dsw^y$.  Due to the strong increase in $\dsw^x$ with $\eta $, 
the average SW stiffness actually grows as the local moments become more non-collinear.  
Similar behavior was found when $t=0$ in Fig.2, where $\dsw^{av}$ was also found to be 
an increasing function of $\eta $.

\begin{figure}
\includegraphics *[scale=0.65]{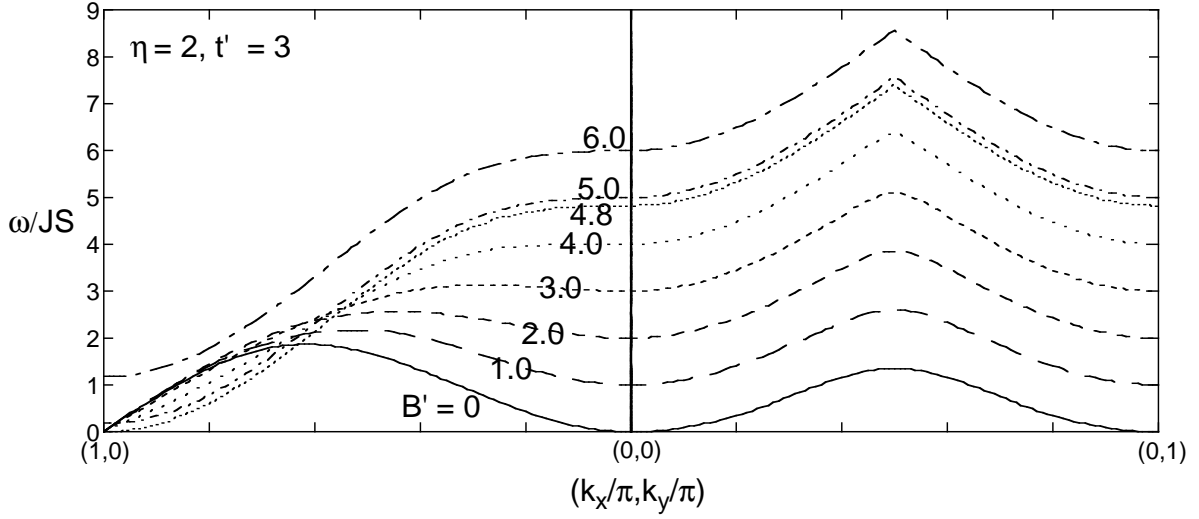}
\caption{
The SW frequencies for $p=0.66$, $\eta=2$, $t'=3$, and various values of $B'$.  
}
\end{figure}

The effect of a magnetic field is shown in Figs.8(b) and 12 for $\eta =2$ and $t'=3$.
Absent from Fig.8(b) is the region of very low fields, where the behavior of the SW stiffness
is complicated by the dependence of the function $\dsw^{\alpha }(k_{\alpha },B)=
(\omega_{\vk }-B)/k_{\alpha}^2$ on the ratio $k_{\alpha }/\sqrt{B'}$ \cite{fis:un}.  
For larger fields, there is a gradual increase in $\dsw^y$ and a decrease in $\dsw^x$ as $B$ 
grows from 0 to $B_c\approx 4.8JS$.  The drop in $\dsw^x$ with field is caused by the 
growth of the SW gap $\omega_0=B$ at $\vk =0$ while the SW frequency continues to vanish 
at $\vQ = (\pi ,0)$ for $B< B_c$, as seen in Fig.12.  Hence, the average SW 
stiffness drops from roughly $1.2JS$ at $B=0$ to $0.54JS$ at $B_c$.  Because a magnetic field 
very quickly destroys the region of phase separation, the crossing point of the $x$ and $y$ stiffnesses 
does not coincide with a region of phase separation.  As expected, the SW stiffnesses are independent of 
field above $B_c$.  Notice from Figs.11 and 12 that the SW velocity at $\vQ $
softens within the CAF phase as $t$ approaches $t_c$ or $B$ approaches $B_c$.  This behavior is 
quite general and occurs at the CAF-FM phase boundary of any magnetic system \cite{rom:00}.  

\section{Discussion and Conclusion}

Applying the results of this work to the manganites requires an estimation of the relevant 
parameters.  Using $S=3/2$, $\eta J S \approx 8.6$ meV, $\eta \sim 1$ \cite{per:97},
and $t\approx 200$ meV for the metallic phase gives $t'\approx 32$ and 
$B'\approx 0.028 B(T)$.  When $\eta =1$, the critical values of the DEV model are $t'_c=5.3$ 
in zero field and $B'_c=2.5$ for $t'=0$.  Hence, the metallic value for the hopping energy 
is more than sufficient to stabilize the FM phase but an applied field of 3 T cannot align the 
local moments when $t'$ is much below $t'_c$.

Metal-insulator transitions in many of the manganites have been successfully interpreted within the 
framework of percolation theory \cite{kim:00,bab:00}.  When the FM fraction $f$ exceeds a
critical value $f_c$, a FM backbone spans the sample and the system becomes metallic.
The percolation threshold $f_c$ in those manganites lies between 0.17 and 0.19.  But it is rather difficult 
to explain the jump in the SW stiffness observed in Pr$_{0.7}$Ca$_{0.3}$MnO$_3$ \cite{fer:02} within 
the framework of percolation theory.  Assuming that the metal-insulator transition in 
Pr$_{1-x}$Ca$_x$MnO$_3$ is produced by the percolation of metallic FM clusters \cite{har:01}, 
there is no reason why the SW stiffness should change when a FM backbone spans the sample.  
Within percolation theory, one would more naturally expect the SW stiffness to 
gradually increase with field as a growing fraction of the sample becomes metallic. 

Estimates of the FM fraction $f$ in Pr$_{1-x}$Ca$_x$MnO$_3$ manganites also pose a difficulty for 
percolation theory.  The accepted range for the percolation threshold $f_c$ within continuum models 
falls between 0.15 and 0.22.  While one experimental estimate of $f_c\approx 0.08$ \cite{har:01}
at the metal-insulator transition lies well below this accepted range, a recent estimate of 
$f_c\approx 0.6$ \cite{mer:03} based on more complete magnetization data lies well above this range.  
At zero field and low temperatures, estimates for $f$ range from 0.3 \cite{mer:03} to 0.5 \cite{rad:01}.  
If those estimates are correct and the FM regions are metallic, then the percolation threshold would 
already be exceeded and the low-temperature phase would be metallic in zero field.

On the other hand, our results clearly indicate that the jump in the SW stiffness at a field of 
3 T cannot be produced by simply aligning the AFM regions while keeping the bandwidth $\sim t$ 
fixed.  As found in Fig.8(b), a magnetic field would act to suppress rather than enhance the average
SW stiffness under those conditions.  Moreover, a field of 3 T corresponding to $B' \approx 0.086 \ll B'_c$ 
will have little affect on the alignment of the local moments.  Whereas experiments indicate 
that the AFM regions begin to shrink under a magnetic field greater than 3 T, a sizeable fraction 
of the sample remains AFM even above the metal-insulator transition \cite{har:01,fer:02,mer:03}.   

The large increase in the SW stiffness predicted by our work as $t'$ increases from 0 to $t'_c$ 
suggests an entirely different scenario:  at the critical field, electrons in the insulating 
and possibly canted FM regions delocalize as the hopping energy $t$ sharply increases.  
Since the integrated optical weight is just $\eke \sim t$, the jump in the 
hopping energy at 3 T should be observable in optical measurements.  Indeed, measurements by 
Okimoto {\it et al.} \cite{oki:99} on Pr$_{0.6}$Ca$_{0.4}$MnO$_3$ reveal a rapid rise 
in $\sigma (\omega )$ and a rapid drop in the CDW gap near the critical field.  If the 
percolation threshold for the FM regions is exceeded when the electrons delocalize, 
as suggested by the measurements discussed above, then the jump in the SW stiffness 
will coincide with the metal-insulator transition.  Otherwise, the metal-insulator 
transition will occur at a slightly larger field.  

Supporting this picture are the plethora of probes that can produce a metal-insulator transition
in Pr$_{1-x}$Ca$_x$MnO$_3$ ($0.3 \le x \le 0.4$).  Besides a magnetic field, application of an 
electric field \cite{asa:97}, high pressure \cite{mor:97}, exposure to x-rays \cite{kir:97}, 
and exposure to visible light \cite{miy:97} all induce a metal-insulator transition.  The identical 
resistivities produced by application of a magnetic field or exposure to x-rays \cite{kir:97} suggest 
a common mechanism:  the excitation of charge carriers out of polaronic traps produced by the 
electron-lattice coupling.  The subsequent relaxation of the lattice \cite{kir:97,rad:01} may 
prevent retrapping of the electrons.

The jump in the SW stiffness as a function of field in Pr$_{0.7}$Ca$_{0.3}$MnO$_3$ 
\cite{fer:02} also bares a striking resemblance to the jump in the SW stiffness observed in
La$_{1-x}$Ca$_x$MnO$_3$ at the metal-insulator transition with $x\approx 0.22 $ \cite{dai:01}.  
In fact, the sizes of the SW stiffnesses on either side of the doping-induced transition in 
La$_{1-x}$Ca$_x$MnO$_3$ are almost exactly the same as on either side of the field-induced 
transition in Pr$_{0.7}$Ca$_{0.3}$MnO$_3$.  This suggests that there are also large increases in the 
hopping energy and electronic kinetic energy in La$_{1-x}$Ca$_x$MnO$_3$ at the
critical concentration of $x\approx 0.22$.  

Compare the jump in the SW stiffness across the metal-insulator transition with the smooth dependence
of the spin-diffusion coefficient $\dsd $ \cite{dai:01,che:03}, which gives the lifetime
$\tau (\vk )=1/\dsd k^2$ of paramagnetic spin fluctuations.  Measurements \cite{dai:01}
of the spin-diffusion coefficient were performed just below $\TC $, where paramagnetic spin 
relaxation occurs within polaronic regions of the FM phase.  While $\dsw $ is a linear 
function of the electronic kinetic energy in the FM phase, $\dsd \chi $ depends only on 
doping and is independent of the electronic bandwidth $W$ in the low-temperature
limit $T \ll W$ \cite{che:03}.  So long as the bandwidth of the polaronic regions 
remains large compared to the temperature, the spin-diffusion coefficient will not change 
across the metal-insulator transition. 

Except in a very narrow range of parameters, phase separation is absent in the DEV model.
Phase separation appears quite commonly in DE models with AFM Heisenberg interactions \cite{dag:01} 
and even occurs near $p=1$ in a DE model without Heisenberg interactions \cite{yun:98a}.  
Results for the DEV model contrast with both a DE model with AFM interactions between all 
neighboring local moments \cite{yun:98b,gol:98,kag:99,gol:00} and a DE model with FM interactions 
in plane but AFM interactions between neighboring planes \cite{aro:98}.  Due to the high symmetry 
of both models, AFM order is not frustrated when $t=0$.  But in both cases, phase separation
occurs before the AFM interactions become strong enough to cant the spins.  Work by 
Golosov \cite{gol:98,gol:00} indicates that the canted phase is destabilized by the presence 
of local degeneracies \cite{inst} that are absent in the DEV model.  Our results
suggest that phase separation only occurs within a very narrow range of parameters when the 
AFM Heisenberg interactions are frustrated and local degeneracies are absent.
For doping concentrations away from multiples of a quarter filling, long-range orbital ordering 
is impossible and magnetic frustration may be present in a wide range of manganites.  
Therefore, the DE model may not provide as straightforward a pathway to phase separation in 
the manganites as believed.  A much more important role may be played by the 
quenched disorder associated with chemical inhomogeneities \cite{bur:01}. 

Rom\'an and Soto \cite{rom:00} pointed out that the nature of the FM and AFM regions 
can be probed by measuring the SW spectrum about the AFM Bragg vector $\vQ $.  In a CAF,
the SW's are not gapped for any field $B$ below $B_c$, as shown in Figs.4 and 12.  By contrast, 
the SW's of a collinear AFM contain a gap at $\vQ$ that grows linearly with field.  So measurements
of the SW spectrum about $\vQ $ may be used to determine whether local regions are canted. 
As discussed elsewhere \cite{fis:un} and seen in the inset to Fig.3(b), the rapid increase
in the SW stiffness for fields near $B^{\star }$ is another signature of a canted phase.
The absence of this behavior in Pr$_{1-x}$Ca$_x$MnO$_3$ for $0.3\le x \le 0.4$ strongly
suggests that the local AFM regions are {\it not} canted. 

Since CE-type AFM ordering occurs at half-filling, the local charge ordering observed 
in the manganites Pr$_{0.6}$Ca$_{0.4}$MnO$_3$ \cite{oki:99,asa:02} and La$_{0.7}$Ca$_{0.3}$MnO$_3$ 
\cite{ada:00,koo:01} would be simplified if the polaronic regions were rich in holes and 
poor in electrons compared to the bulk.  Our model provides a natural explanation
for this behavior, since the electronic fraction on $b$ sites is substantially smaller than the
fraction on $a$ sites as the electrons avoid regions with more pronounced AFM order.  

This paper has studied the general effects of AFM interactions and non-collinearity on the
magnetic ordering and spin dynamics of DE systems.  The competition between DE and AFM interactions 
is responsible for several interesting properties.  In a narrow region of hopping energies, 
weak phase separation occurs as the FS topology changes from closed to open.  Because electrons 
prefer to occupy sites that are coupled by FM interactions, a CDW appears in the absence of
a CDW gap.  For finite Hund's coupling, the electron spins are more closely aligned than the 
local moments of the CAF.  The CAF becomes unstable above critical values of field 
and hopping energy and below a critical value of $\eta $.  Perhaps the most surprising result  
is that DE changes none of the qualitative features of the CAF state.  The SW gap $\omega_0$ at 
$\vk =0$ remains equal to $B$ and the gap $\omega_{\vQ}$ at the AFM Bragg vector $\vQ $ continues 
to vanish for any hopping energy below $t_c$.   The average SW stiffness still softens with 
magnetic field and hardens with strengthening AFM interactions; the SW velocity at $\vQ$ still 
vanishes at the CAF-FM transition.  Clearly, a great deal can be learned about the general 
properties of CAF's, even of the more complex variations that appear in many manganites, by  
considering simple models such as the DEV model.

It is a pleasure to acknowledge helpful conversations with Drs A Chernyshev, P-C Dai, E Dagotto,
J Fernandez-Baca, D Golosov, M Katsnelson, N Furukawa, W Saslow, and A Zheludev.  This research 
was sponsored by the US Department of Energy under contract DE-AC05-00OR22725 with Oak Ridge 
National Laboratory, managed by UT-Battelle, LLC.


\begin{references}

\bibitem{dag:01} For an overview of phase separation in the manganites,
see Moreo A, Yunoki S, and Dagotto E 1999 {\it Science} {\bf 283} 2034
and Dagotto E, Hotta T, and Moreo A 2001 {\it Phys. Rep.} {\bf 233} 1.

\bibitem{ada:00} Adams C P, Lynn J W, Mukovskii Y M, Arsenov A A, and Shulyatev D
2000 {\it Phys. Rev. Lett.} {\bf 85}, 3954.

\bibitem{koo:01} Koo T Y, Kiryukhin B, Sharma P A, Hill J P, and Cheong S-W 2001
{\it Phys. Rev. B} {\bf 64} 220405.

\bibitem{che:03} Chernyshev A and Fishman R S 2003 {\it Phys. Rev. Lett.}
{\bf 90} 177202. 

\bibitem{jir:85} Jirak Z, Krupicka S, Simsa Z, Dlouha M, and Vratislav S,
1985 {\it J. Magn. Magn. Mater.} {\bf 53} 153.

\bibitem{yos:95} Yoshizawa H, Kawano H, Tomioka Y, and Tokura Y 1996 {\it Phys. Rev. B}
{\bf 52} R13145; 1996 {\it J. Phys. Soc. Japan} {\bf 65} 1043.

\bibitem{oki:99} Okimoto Y, Tomioka Y, Onose Y, Otsuka Y, and Tokura Y 1999 {\it Phys. Rev. B}
{\bf 59} 7401.

\bibitem{deac:01} Deac I G, Mitchell J F, and Schiffer P 2001 {\it Phys. Rev. B} {\bf 63}
172408.

\bibitem{rad:01} Radaelli P G, Ibberson R M, Argyriou D N, Casalta H, Anderson K H, 
Cheong S-W, and Mitchell J F 2001 {\it Phys. Rev. B} {\bf 63} 172419.

\bibitem{har:01} Hardy V, Wahl A, and Martin C 2001 {\it Phys. Rev. B} {\bf 64} 064402.

\bibitem{fer:02} Fernandez-Baca J A, Dai P-C, Kawano-Furukawa H, Yoshizawa H, 
Plummer E W, Katano S, Tomioka Y, and Tokura Y 2002 {\it Phys. Rev. B}
{\bf 66} 054434.

\bibitem{sim:02} Simon Ch, Mercone S, Guiblin N, Martin C, Br\^ulet A, and Andr\'e G 2002 
{\it Phys. Rev. Lett.} {\bf 89} 207202.

\bibitem{mer:03} Mercone S, Hardy V, Martin C, Simon C, Saurel D, and Br\^ulet A 2003
{\it Phys. Rev. B} {\bf 68} 094422.

\bibitem{and:55} Anderson P W and Hasegawa H 1995 {\it Phys. Rev.} {\bf 100} 675.

\bibitem{deg:60} DeGennes P-G 1960 {\it Phys. Rev.} {\bf 118} 141.

\bibitem{vil:77} Villain J 1977 {\it J. Phys. C} {\bf 10} 1717.

\bibitem{ber:86} Berge B, Diep H T, Ghazali A, and Lallemand P 1986 {\it Phys. Rev. B}
{\bf 34} 3177.

\bibitem{gab:89} Gabay M, Garel T, Parker G N, and Saslow W M 1989 {\it Phys. Rev. B}
{\bf 40} 264.

\bibitem{sas:92} Saslow W M and Erwin R 1992 {\it Phys. Rev. B} {\bf 45} 4759.

\bibitem{yun:98b} Yunoki S and Moreo A 1998 {\it Phys. Rev. B} {\bf 58} 6403. 

\bibitem{gol:98} Golosov D I, Norman M R, and Levin K 1998 {\it Phys. Rev. B} {\bf 58}
8617.

\bibitem{kag:99} Kagan M Yu, Khomskii D I, and Mostovoy M V 1999 {\it Eur. Phys. J. B}
{\bf 12} 217.

\bibitem{gol:00} Golosov D I 2000 {\it Phys. Rev. Lett.} {\bf 84} 3974; 2000 {\it
J. Appl. Phys.} {\bf 87} 5804; 2002 {\it ibid.} {\bf 91} 7508.

\bibitem{inst}  In the canted phase of a two-sublattice model with AFM interactions, the 
local moment on any site of sublattice $a$ must make an angle $\Theta $ with the local moments 
on sublattice $b$.  So the possible orientations for any $a$ spin trace out a cone around the
equilibrium direction of the $b$ spins \cite{gol:98,gol:00}.  In the canted phase of a 
two-sublattice model with FM interactions in plane but AFM interactions between planes, 
the moments in any plane $a$ are free to rotate about the equilibrium direction of the 
moments on neighboring planes $b$.  These local or planar degeneracies are absent in 
the DEV model.

\bibitem{miz:00} Mizokawa T, Khomskii D I, and Sawatzky G A 2000 {\it Phys. Rev. B}
{\bf 63} 024403.

\bibitem{pol} While the FM coupling between neighboring Mn$^{4+}$ and Mn$^{3+}$ ions are induced by the 
hopping of electrons those sites, that coupling is significantly enhanced by the polaronic distortions 
of the O atoms around some Mn$^{4+}$ ions \cite{miz:00}.  So the Heisenberg coupling $J$ incorporates the 
enhancement of the FM interaction beyond that produced by the uniform hopping of electrons between 
neighboring sites on an undistorted lattice. 

\bibitem{fis:04} Fishman R S 2004 {\it Phys. Rev. B} ({\it in press}).

\bibitem{wal:80} Walker L R and Walstedt R E 1980 {\it Phys. Rev. B} {\bf 22} 3816.

\bibitem{fis:un} Fishman R S ({\it unpublished}).

\bibitem{fbz}  Since there are four inequivalent sites on the Villain lattice, it might 
seem that the first Brillouin zone should be reduced in both the $x$ and $y$ directions.  
However, all $a$ or $b$ sites experience the same environment regardless of whether the 
spins tilt to the $+x$ or $-x$ directions:  each $a$ spin, for example, makes an angle of 
$\tb -\ta $ with its $b$ neighbors and an angle $2\ta $ with its $a$ neighbors.

\bibitem{rom:00} Rom\'an  J M and Soto J 2000 {\it Phys. Rev. B} {\bf 62} 3300.

\bibitem{asa:02} Asaka T, Yamada S, Tsutsumi S, Tsurata C, Kimoto K, Arima T, and  
Matsui Y 2002 {\it Phys. Rev. Lett.} {\bf 88} 097201.

\bibitem{hal:00} Halboth C J and Metzner W 2000 {\it Phys. Rev. Lett.} {\bf 85} 5162.

\bibitem{val:01} Valenzuela B and Volzmediano M A H 2001 {\it Phys. Rev. B} {\bf 63} 153103.

\bibitem{fur:95} Furukawa N 1995 {\it J. Phys. Soc. Japan} {\bf 64} 2754;
1999 {\it Physics of Manganites} ed T A Kaplan and S D Mahanti (New York: Plenum) p 1.

\bibitem{per:97} Perring T G, Aeppli G, Moritomo Y, and Tokura Y 1997 {\it Phys. Rev. Lett.}
{\bf 78} 3197.

\bibitem{kim:00} Kim K H, Uehara M, Hess C, Sharma P A, and Cheong S-W 2000 {\it Phys.
Rev. Lett.} {\bf 84} 2961.

\bibitem{bab:00} Babushkina N A, Taldenkov A N, Belova L M, Chistotina E A, Gorbenko O Yu,
Kaul A R, Kugel K I, and Khomskii D I 2000 {\it Phys. Rev. B} {\bf 62} R6081.

\bibitem{asa:97} Asamitsu A, Tomioka Y, Kuwahara H, and Tokura Y 1997 {\it Nature} {\bf 388}
50.

\bibitem{mor:97} Moritomo Y, Kuwahara H, Tomioka Y, and Tokura Y 1997 {\it Phys. Rev. B}
{\bf 55} 7540.

\bibitem{kir:97} Kiryukhin V, Casa D, Hill J P, Keimer B, Vigliante A, Tomioka Y,
and Tokura Y 1997 {\it Nature} {\bf 386} 813.

\bibitem{miy:97} Miyano K, Tanaka T, Tomioka Y, and Tokura Y 1997 {\it Phys. Rev. Lett.}
{\bf 78} 4257.

\bibitem{dai:01} Dai P-C, Fernandez-Baca J A, Plummer E W, Tomioka Y, and Tokura Y 2001 
{\it Phys. Rev. B} {\bf 64} 224429.

\bibitem{yun:98a} Yunoki S, Hu J, Malvezzi A L, Moreo A, Furukawa N, and Dagotto E 1998
{\it Phys. Rev. Lett.} {\bf 80} 845.

\bibitem{aro:98} Arovas D P and Guinea F 1998 {\it Phys. Rev. B} {\bf 58} 9150.

\bibitem{bur:01} Burgy J, Mayr M, Martin-Mayor V, Moreo A, and Dagotto E 2001 {\it Phys.
Rev. Lett.} {\bf 87} 277202.

\end{references}
\end{document}